\documentclass[a4paper,11pt, titlepage, twoside]{article}

\usepackage{titling}
\usepackage{amssymb}
\usepackage{diagbox}
\usepackage{rotating}
\usepackage{float}

\title{Deliverable Report:}
\author{Alexander Novikov (DTU),
Luis A. Garcia-Reyes (UPC),
Anup Joshi (EC Lille),
Ahmar Farooque (U Porto),
Junaid Khalid (U Cardiff)\\
Reviewed by: Dirk Van Hertem (KUL)}
\date{July 2024}

\usepackage{lipsum}
\usepackage[T1]{fontenc}
\usepackage{parskip}
\usepackage[sorting = none, backend = bibtex, style=numeric-comp]{biblatex}
\usepackage[a4paper,top=3cm,bottom=4cm,left=2.5cm,right=2.5cm]{geometry}

\usepackage{graphicx} 
\usepackage{float}
\usepackage{eurosym}
\usepackage{amsmath}
\usepackage{textgreek}
\usepackage{steinmetz}
\usepackage{systeme}
\usepackage{sectsty}
\usepackage{multirow}
\usepackage{caption}
\usepackage{subcaption}
\usepackage{wrapfig}
\usepackage{listings}
\usepackage[colorlinks=true, allcolors=blue]{hyperref}
\usepackage{color} 
\usepackage[dvipsnames]{xcolor}
\usepackage[section]{placeins} 
\usepackage{longtable}
\usepackage{mathtools}
\usepackage{commath}
\usepackage{dtk-logos}
\usepackage{fancyhdr}
\usepackage{titlesec}
\usepackage{transparent}
\usepackage{booktabs}
\usepackage{graphicx}
\usepackage[sorting = none, backend = bibtex, style=numeric-comp]{biblatex}
\usepackage{acronym}

\usepackage{listings}
\definecolor{codegreen}{rgb}{0,0.6,0}
\definecolor{codegray}{rgb}{0.5,0.5,0.5}
\definecolor{codepurple}{rgb}{0.58,0,0.82}
\definecolor{light-gray}{gray}{0.95}

\lstdefinestyle{Python}{
  backgroundcolor=\color{light-gray}, 
  commentstyle=\color{codegreen},
  keywordstyle=\color{violet},
  numberstyle=\tiny\color{codegray},
  stringstyle=\color{codepurple},
  basicstyle=\ttfamily\footnotesize,
  emph={self,__init__},          
  emphstyle=\color{red},
  breakatwhitespace=false,         
  breaklines=true,                 
  captionpos=b,                    
  keepspaces=true,                 
  numbers=left,                    
  numbersep=5pt,                  
  showspaces=false,                
  showstringspaces=false,
  showtabs=false,                  
  tabsize=2
}

\lstdefinestyle{Matlab}{
    backgroundcolor=\color{light-gray}, 
    commentstyle=\color{codegreen},
    breaklines=true,%
    morekeywords={matlab2tikz},
    basicstyle=\ttfamily\footnotesize,%
    keywordstyle=\color{blue},%
    morekeywords=[2]{1}, keywordstyle=[2]{\color{black}},
    identifierstyle=\color{black},%
    stringstyle=\color{codepurple},
    showstringspaces=false,
    numbers=left,%
    numberstyle={\tiny \color{black}},
    numbersep=5pt, 
    emph=[1]{for,end,break},emphstyle=[1]\color{blue}, 
}

\chapterfont{\color{Blue}}  
\sectionfont{\color{Blue}}  
\subsectionfont{\color{RoyalBlue}}

\newcommand{\sectionbreak}{\clearpage}

\begin{document}

\begin{titlepage}

 \includegraphics[height=2 cm,keepaspectratio]{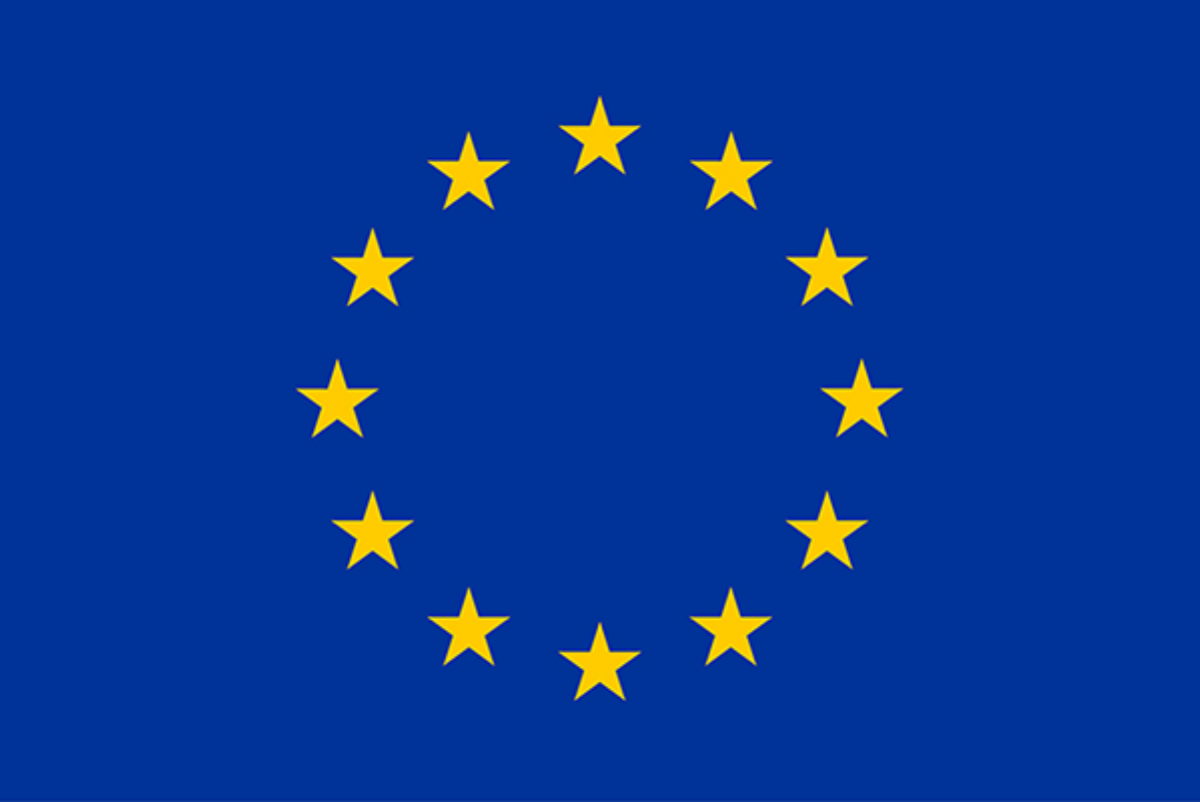}  
 \raggedleft
     \includegraphics[height=2cm,keepaspectratio]{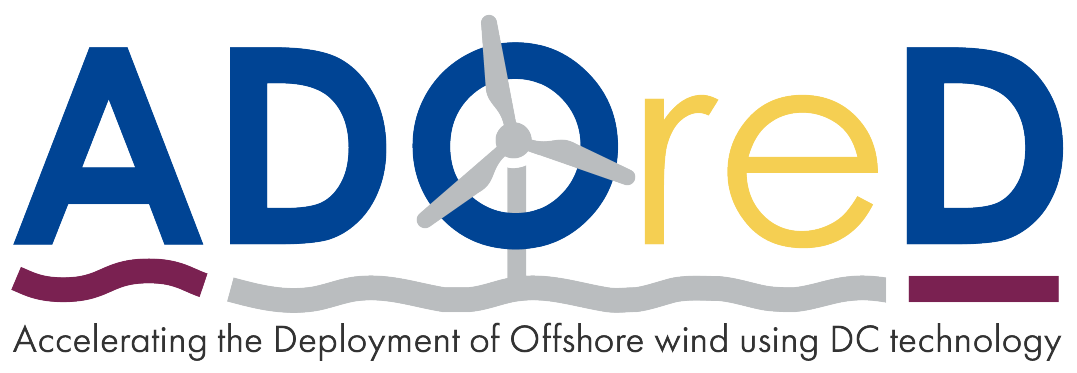}

 \vfill
    \hspace{2mm}
\centering
    {\huge \textbf{ADOreD:}}\\
    \vspace{5mm}
    {\Large \textbf{Accelerating the Deployment of Offshore wind using DC technology}}\\
    \Large \textbf{Grant Agreement number : 101073554}\\
    \vspace{15mm}
    \Huge \textcolor{Blue}{\textbf{\thetitle}}\\
    \vspace{10mm}
    \Large\textbf{Protection challenges and solutions on power systems towards 100\% inverter-based resources and energy storage systems}\\
    \vspace{10mm}
    \Large\today\\  
    
    \vspace{15mm}
    \hspace{2mm}
    \theauthor
\begin{table}[H]
\centering
\begin{tabular}{|l|l|}
\hline
Deliverable number     & D3.1               \\ \hline
Work Package           & WP3: AC/DC interaction    \\ \hline
Dissemination level    & Sensitive             \\ \hline
Project period         & Month 15 \\ \hline
Due date               & Month 12   \\ \hline
Actual submission date & July 1, 2024  \\ \hline
\end{tabular}
\end{table}

    \end{titlepage}

\tableofcontents
\listoffigures
\listoftables
\section*{Acronyms}
\begin{acronym}[YesIcan]
      \acro{wt}[WT]{\acroextra{\dotfill} Wind Turbine}
      \acro{owt}[OWT]{\acroextra{\dotfill} Offshore Wind Turbine}
      \acro{hvdc}[HVDC]{\acroextra{\dotfill} High Voltage Direct Current}
      \acro{ac}[AC]{\acroextra{\dotfill} Alternating Current}
      \acro{dc}[DC]{\acroextra{\dotfill} Direct Current}
      \acro{mw}[MW]{\acroextra{\dotfill} Mega Watts}
      \acro{wecs}[WECS]{\acroextra{\dotfill} Wind Energy Conversion Systems}      
      \acro{pv}[PV]{\acroextra{\dotfill} PhotoVoltaics}
      \acro{sg}[SG]{\acroextra{\dotfill} Synchronous Generator}
     \acro{lvrt}[LVRT]{\acroextra{\dotfill} Low Voltage Ride Through}
     \acro{gfm}[GFM]{\acroextra{\dotfill}  Grid Forming Converter}
     \acro{gfl}[GFL]{\acroextra{\dotfill} Grid Following Converter}
      \acro{ffci}[FFCI]{\acroextra{\dotfill }Fast Fault Current Injection}
     \acro{rocof}[ROCOF]{\acroextra{\dotfill} Rate of change of frequency}
     \acro{csa}[CSA]{\acroextra{\dotfill} Current Saturation Algorithm}
     \acro{avr}[AVR]{\acroextra{\dotfill} Automatic Voltage Regulator}
     \acro{pll}[PLL]{\acroextra{\dotfill} Phase Locked Loop}
      \acro{cct}[CCT]{\acroextra{\dotfill} Critical Clearing Time}
       \acro{cca}[CCA]{\acroextra{\dotfill} Critical Clearing Angle}
      \acro{vi}[VI]{\acroextra{\dotfill} Virtual Impedance}
      \acro{ibr}[IBR]{\acroextra{\dotfill} Inverter-Based Resources}
      \acro{res}[RES]{\acroextra{\dotfill} Renewable Energy Sources}
      \acro{ied}[IED]{\acroextra{\dotfill} Intelligent Electronic Devices}
      \acro{scr}[SCR]{\acroextra{\dotfill} Short-Circuit Ratio}
      \acro{lom}[LOM]{\acroextra{\dotfill} Loss Of Main}
      \acro{fft}[FFT]{\acroextra{\dotfill} Fast Fourier Transform}
      \acro{pcc}[PCC]{\acroextra{\dotfill} Point of Common Coupling}
      \acro{ai}[AI]{\acroextra{\dotfill}  Artificial Intelligence}
      \acro{ml}[ML]{\acroextra{\dotfill} Machine Learning}
\end{acronym}

\section{Introduction}
\label{sec:intro}
In compliance with global climate action under the Paris Agreement \cite{paris_agreement}, the EU is implementing the European Green Deal \cite{cellar} and aiming for climate neutrality by 2050. One of the key areas of future development is the decarbonization of the energy sector. Renewable energy sources  (RES) are considered a key technology that will help achieve the aforementioned climate goals. RES based power systems differ from traditional system due to the reliance on power electronic equipment to convert the renewable energy into electricity. The increasing share of renewable sources in power systems will generate unprecedented challenges that could affect their reliability and resilience.

The development of the energy system, driven by the continued scale of renewable energy technologies (wind turbines (WT), offshore wind farms (OWT), battery storage facilities, and photovoltaic panels) and their integration, requires fundamentally new approaches. As part of the future transition, a new power system will emerge that exists of a low-level layer, much like our current AC system, and upper-level layer or backbone system that will consist of a High Voltage Direct Current (HVDC) grid \cite{VANHERTEM2016hvdc,VANHERTEM2010}. Protection against faults, stability, and interaction between different components in these extensive combined AC and DC grids require a new understanding of the power system principles. Additionally, new power systems models and simulation tools are needed to investigate these principles. As the transition to a future power system that is 100\% based on power converters complicates these challenges, it becomes crucial to analyze the integration of renewable energy sources into zero-emission systems. To address these challenges, innovative solutions are needed for the safety and interactions between AC and DC networks before DC networks can be more widely installed \cite{adored}.

In power systems, protection systems detect, locate, and isolate faults (short-circuit paths) to protect assets while allowing the remaining system to continue operating. Protection is traditionally based on the detection of the flow of large fault currents generated by synchronous machines towards the fault location. This substantial overcurrent is measured by sensors and protection relays use advanced algorithms to determine whether the measured signal is i) a fault; and ii) within their protection zone (and to which they need to act) \cite{bookprotection}.

However, in modern power systems, converter controls typically limit the contribution of fault current to a value close to the converter's rated value \cite{orbitjundi, mukeshpes}. Due to the fact that fault current contribution is typically limited to 120\% of its rated currents, relays that differentiate between normal overcurrents and faults based on short-circuit current levels may operate incorrectly. Additionally, the fault current contribution from inverter-based resources (IBR) may not contain enough negative and zero sequence currents for the proper operation of directional relays \cite{Joule}. Thirdly, converters support the grid voltage to a fixed value within the reach of the converter, while beyond the reach of the converter, the voltage drops quickly.  Consequently, the existing protection relay system may not be suitable to meet the expected accuracy and reliability.

This report reviews existing issues highlighted in recent years in academic literature and research articles related to the protection challenges in power systems towards 100 \% inverter-based resources also denominated as converter-based resources in next chapters. The structure of the report is as follows:
The specific characteristics of the renewable energy systems (RES) are described in the first chapter and the second chapter provides an overview of challenges in protection of intelligent electronic devices (IED) in the power system. The third chapter outlines challenges in converter control during a fault, and the final chapter presents the main solutions suggested in research papers over the last decade.

\section{Special aspects of 100\% RES based systems}
\label{sec:special_aspects}

A 100\% Renewable Energy Sources (RES) power system differs significantly in design and operation from a conventional power system, both in terms of implementation and functionality. Considering the development challenges and high investment costs associated with these systems, many countries worldwide are on the path towards decarbonizing their electrical systems. In Europe, from 2021 to 2022, there was an increase in installed capacity of 53\,969 MW from renewable energy sources, marking a considerable rise compared to the previous year's increase \cite{irena}. However, along with the increase in capacity, various challenges have arisen for the massive integration of these renewable sources. Among the primary challenges are the variability of generation sources and the inverter-based interfaces that allow connection to the grid of these sources. Transmission and protection systems must be redesigned to improve and enable the reliable operation of modern power systems. Solar and wind energy are the main focus of interest as they are expected to be the prevailing RES, which will bring about a change in the way the electricity market operates. Table \ref{table_aspects} provides a comparison of different research articles addressing various issues and key aspects within 100\% RES-based systems.

\begin{table}[h]
\caption{Main features of 100\% RES based systems}
\label{table_aspects}
\centering
\begin{tabular}{lcccccc}
\hline
\backslashbox{Subject}{Reference} & \textbf{\cite{Joule}} & \textbf{\cite{adaptative100}} & \textbf{\cite{protectionschemes100}} & \textbf{\cite{operational100}} & \textbf{\cite{addressingtech}} & \textbf{\cite{citcea}}  \\ \hline
Definitions about 100\% RES  &  $\checkmark$ & $\checkmark$ & $\checkmark$ & $\checkmark$ &    $\checkmark$ &    $\checkmark$ \\
Economical balance (supply-demand)  &  $\checkmark$ & $\times$ & $\times$ & $\times$ &    $\times$ &    $\times$ \\
RES Curtailment and peaking &  $\checkmark$ & $\times$ & $\times$ & $\times$ &    $\checkmark$ &    $\times$ \\
 Seasonal storage &  $\checkmark$ & $\times$ & $\times$ & $\times$  &  $\checkmark$  &    $\times$ \\
Inertia reduction  &  $\times$ & $\times$ & $\checkmark$ & $\times$ &    $\checkmark$  &    $\checkmark$ \\
 Ancillary services &  $\times$ & $\checkmark$ & $\checkmark$ & $\checkmark$ &    $\checkmark$  &  $\times$ \\
Markets adaptability to RES  &  $\checkmark$ & $\times$  & $\times$ & $\times$ &    $\checkmark$ &  $\times$ \\
Converter control  &  $\times$ & $\checkmark$ & $\checkmark$ & $\checkmark$ &    $\checkmark$ &    $\checkmark$ \\
Converter limitations  &  $\times$ & $\times$ & $\times$ & $\times$ &    $\checkmark$ &    $\checkmark$ \\
Voltage and frequency stability &  $\times$ & $\times$ & $\checkmark$ & $\checkmark$ &    $\checkmark$  &    $\checkmark$ \\
Rotor and angle stability  &  $\times$ & $\times$  & $\times$ & $\checkmark$ &    $\times$ &    $\checkmark$ \\
Protection in 100\% RES  &  $\times$ & $\checkmark$ & $\checkmark$ & $\checkmark$ &   $\checkmark$  &  $\times$ \\
 \hline
\end{tabular}
\end{table}

Effectively addressing these unique aspects of 100\% RES necessitates a comprehensive and collaborative approach involving  industry stakeholders (both system operators and technology providers), research institutions, governmental agencies (including government, regulators and policy makers), and communities to expedite the transition towards a sustainable energy future. Then, the specific aspects of 100\% RES can be divided into the following key categories:
\begin{enumerate}
    \item Grid Integration.
    \item Power electronic power conversion.
\end{enumerate}

\subsection{Grid Integration}

The incorporation of substantial levels of renewable energy into the grid mandates substantial enhancements to transmission and distribution infrastructure due to the specific characteristics of operation of the renewables, which are:
\begin{enumerate}
    \item Intermittency and variability.
    \item Energy storage.
    \item Diverse renewable resources.
    \item Descentralization and distributed generation.
    \item Community engagement and public acceptance.
    \item Technological innovation.
\end{enumerate}

\subsubsection*{Intermittency and Variability}

Renewable energy sources, like solar and wind, exhibit an intermittent and variable behavior, contingent upon weather conditions and time of day. Effectively managing this variability necessitates advanced forecasting techniques, energy storage solutions, and adaptable grid infrastructure to ensure a consistent power supply to end users \cite{IPCC}.

\subsubsection*{Energy Storage}

Vital for balancing supply and demand within renewable energy systems, energy storage technologies, including batteries, pumped hydro storage, and thermal energy storage, play a pivotal role. Implementing efficient and cost-effective storage solutions is imperative to sustain uninterrupted power during periods of low renewable generation \cite{futuresupply}.

\subsubsection*{Diverse Renewable Resources}

To mitigate the challenges posed by intermittency, a diverse array of renewable energy sources is often necessary. Harnessing a mix of solar, wind, hydroelectric, geothermal, and biomass resources can bolster a more dependable and resilient energy supply \cite{irena}.

\subsubsection*{Decentralization and Distributed Generation}

Renewable energy systems frequently entail decentralized generation, wherein power is produced in closer proximity to consumption points. Technologies facilitating distributed generation, such as rooftop solar panels and small wind turbines, empower consumers to generate their electricity and diminish dependence on centralized power plants \cite{EUagency}.

\subsubsection*{Community Engagement and Public Acceptance}

Garnering public support for renewable energy projects is paramount to their successful implementation. Engaging with local communities, addressing concerns related to visual impacts, noise, and land use, and ensuring equitable distribution of economic benefits can cultivate acceptance of renewable energy developments \cite{socialacceptance}.

\subsubsection*{Technological Innovation}

Continuous research and development in renewable energy technologies is indispensable for cost reduction, efficiency enhancement, and overcoming technical impediments. Progress in areas such as solar photovoltaics, wind turbine design, energy storage, and grid integration is essential for advancing the deployment of renewable energy on a global scale \cite{IEA}.

\subsection{Power electronic power conversion}

Power electronic converters are crucial for integrating renewable energy sources like batteries, PVs, WTs, and HVDC systems into the grid. These converters enable precise control over the power flow and ensure compatibility between different types of energy sources and the grid. Advanced control strategies for power electronic converters are essential to manage their behavior and interaction with the grid, ensuring stability and efficiency in 100\% RES systems. In these systems the main challenges associated to the power electronics conversion are:
\begin{itemize}
    \item Converter control and limitations.
    \item Voltage and frequency stability.
    \item Protection systems.
    \item Low inertia.
    \item Low system strength. 
\end{itemize}

\subsubsection*{Converter control and limitations}

Effective control of power electronic converters is vital for maintaining grid stability and reliability. These converters must handle rapid changes in power demand and supply, provide voltage and frequency regulation, and ensure the seamless integration of various renewable energy sources. However, limitations in converter technology, such as efficiency losses and thermal management challenges, must be addressed through ongoing research and development to enhance their performance and reliability \cite{Joule, futuresupply}.

\subsubsection*{Voltage and frequency stability}

Maintaining voltage and frequency stability in a power system dominated by renewable energy sources is challenging due to the variable nature of these sources. Power electronic converters play a key role in stabilizing voltage and frequency by providing fast response and control capabilities. Innovative control algorithms and advanced converter designs are necessary to ensure stable operation in 100\% RES systems \cite{Joule, futuresupply}.

\subsubsection*{Protection systems}

Protecting the power grid in a 100\% RES scenario involves developing new protection schemes that can rapidly detect and isolate faults while accommodating the unique characteristics of renewable energy sources and power electronic converters. Traditional protection systems must be adapted to handle the fast dynamics and lower inertia of renewable-based power systems, ensuring reliable and safe operation \cite{adaptative100}.

\subsubsection*{Low inertia}

Traditional power plants based on synchronous generators are characterized by high inertia, which reduces the immediate effect of power imbalances (e.g.\, due to outage) on the system frequency. The stored energy in the rotor of the synchronous machines is used to counteract changes in the system frequency. A high level of power grid equivalent inertia leads to lower rate of change of frequency (ROCOF). Since power electronic converters do not have any rotating mass, nor any other substantial energy storage,  the contribution of power electronic converters to the inertia is very small, yielding to very low inertia levels  in converter-dominated power system. This leads to the number of problems in power system in case of disturbances:

\begin{enumerate}
\item Acceleration of transient phenomenon (faster system dynamics).
\item High ROCOF and frequency deviations \cite{6}.
\item Complication of frequency regulation in the system \cite{7}.
\item Maloperation of ROCOF-based protection for Loss of Main (LoM) \cite{7}.
\item Grid-Synchronization Stability of Converter \cite{8, 9}.

\end{enumerate}

\subsubsection*{Low system strength}

System strength is traditionally defined with the help of short circuit ratio (SCR) value. Although, definition and evaluation of power system strength is widely debated topic nowadays \cite{debates, Gordon}.

According to \cite{10,SCR3,SCR2}, the system strength in traditional power systems can be defined by short-circuit ratio (SCR):

\begin{equation}
    SCR = \frac{\text{Three-phase-short-circuit-MVA-at-the-PCC}}{\text{Active-power-rating-of-the-DC-system}}
\end{equation}

While the $Three-phase-short-circuit-MVA-at-the-PCC$ also known as Short Circuit Power of AC grid as seen from the PCC may be calculated by:

\begin{equation}
    \text{Three-phase-short-circuit-MVA-at-the-PCC} = \frac{U_{ac}^2}{Z_{th}}
\end{equation}

where $U_{ac}$ is the AC system nominal voltage and $Z_{th}$ is the Th\'evenin equivalent impedance as seen from the PCC. 

\begin{figure}
    \centering
    \includegraphics[width=0.5\linewidth]{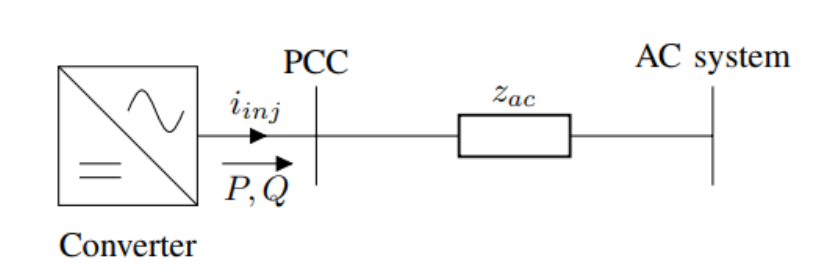}
    \caption{Equivalent circuit of the AC system as seen from the PCC \cite{SCR2}}
    \label{fig_SCR}
\end{figure}

A higher value of the Th\'evenin equivalent impedance of AC system corresponds to a weaker system strength and the greater voltage drops in case of disturbance as shown in Figure \ref{fig_SCR}.

In a power system primarily based on synchronous generators, the generators act as voltage sources behind an impedance during faults, and this principle holds true. However, in a power system dominated by converters, the fault response of a converter is mainly dictated by its controllers and can be most accurately represented as behaving like a controlled current source. The short-circuit contribution from converters is not significantly correlated to the actual impedance of the converters and the system when the converter thermal limit is reached. Moreover, in the transmission level the short-circuit contribution from converters tends to be saturated and less than that of synchronous generators due to the constrained thermal capacity of the semiconductors involved.\cite{SCR2} Although the different scenarios can be noticed on distribution levels where significant amounts of distributed sources may provide high fault currents as reported in \cite{Gordon2}.

Short Circuit Power may be calculated based on the formula below in this case:

\begin{equation}
    \text{Three-phase-short-circuit-MVA-at-the-PCC} = \sqrt{3} \times {U_{ac}} \times \text{SC-current-in-kA}
\end{equation}

where $SC-current-in-kA$ is obtained from simulation which includes response of converter-based sources control and semiconductor thermal limitations.

Massive integration of converter-based sources may decrease fault current levels in the system and consequently decreases the system strength. Converters do not have a meaningful short‐term overcurrent rating since overcurrents in semiconductors lead to damaging temperature rises very quickly because of the very low thermal mass.

Low system strength can have negative impacts on the grid:

\begin{enumerate}
\item Low fault current level.
\item Wider range of voltage variations.
\item More harmonics.
\item Impact on protection systems \cite{11}.
\item Impact on inverter-driven stability \cite{12}.
\item Impact on voltage stability.
\item Impact on power quality.

\end{enumerate}

\subsection{Types of converter-based energy sources}

Converter-based energy sources refer to renewable energy systems that utilize power electronic converters to convert the variable output of renewable sources into grid-compatible electricity. Here are some types of converter-based energy sources:

\subsubsection*{Solar Photovoltaic (PV) Systems}

Solar PV systems convert sunlight directly into electricity using photovoltaic cells. Power electronic converters, such as inverters, are used to convert the DC output of solar panels into AC electricity suitable for the grid \cite{pv}.

\subsubsection*{Wind Energy Conversion Systems (WECS)}

Wind turbines generate electricity from the kinetic energy of wind. Power electronic converters, typically in the form of wind turbine converters and grid-side converters, are used to convert the variable AC output of wind turbines into stable AC electricity for the grid \cite{icrera}.According with the standard IEC-61400-27-1, there are 4 different types of WTs \cite{DTUwind, WTdynamics}: 
\begin{description}
    \item[Type I] This configuration denotes the fixed speed WT with an asynchronous induction generator (IG) directly connected to the grid via a transformer. 
    \item[Type II] This configuration corresponds to a limited variable speed wind turbine with a variable rotor resistance.
    \item[Type III] In this configuration, there is a doubly fed induction generator (DFIG) with its stator connected directly to the grid and its rotor is connected through a back-to-back converter. Here the speed range is within the synchronous speed.
    \item[Type IV] In this type, the the variable speed WT has a generator that is connected to the grid through a frequency converter. This WT type configuration, use either synchronous or asynchronous generators.
\end{description}

\begin{figure}[h]
\centering
\subfloat[Type I]{\includegraphics[scale=0.35]{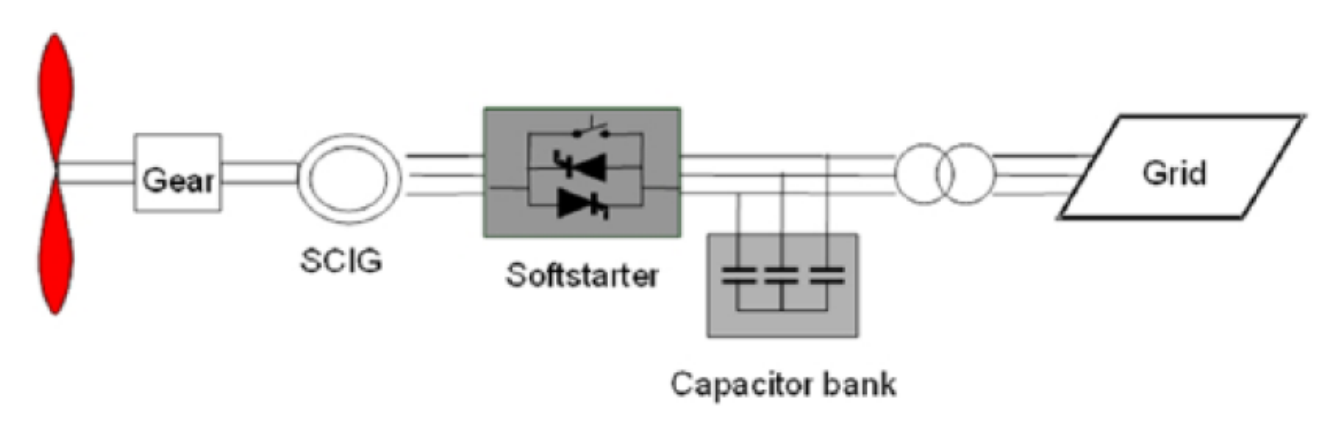}%
\label{fig_first_case}}
\subfloat[Type II]{\includegraphics[scale=0.35]{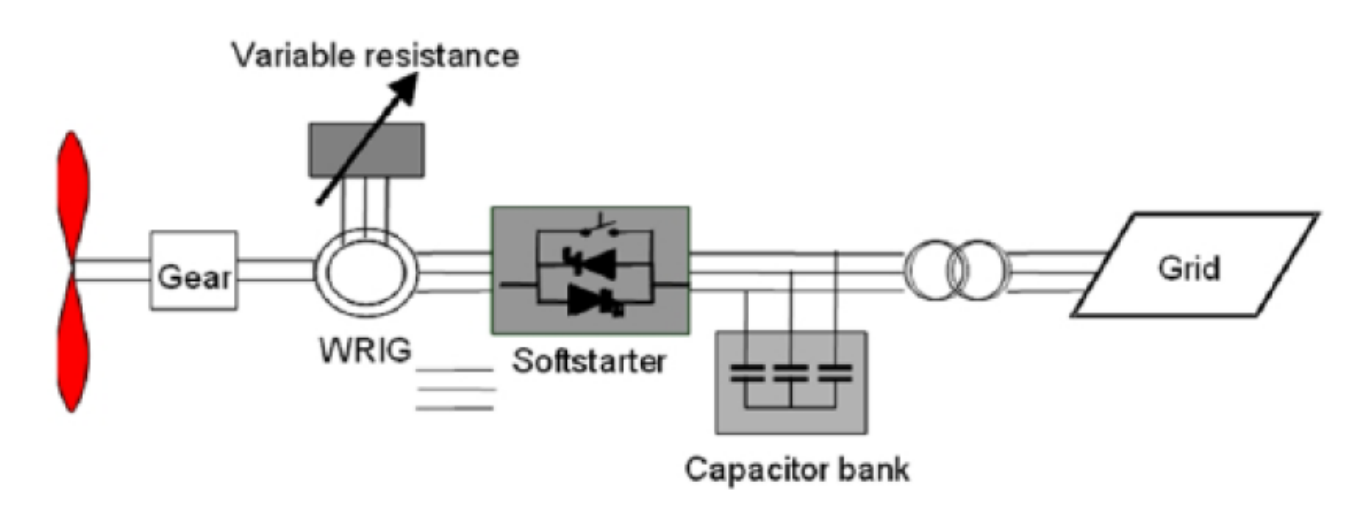}%
\label{fig_second_case}} \\
\subfloat[Type III]{\includegraphics[scale=0.35]{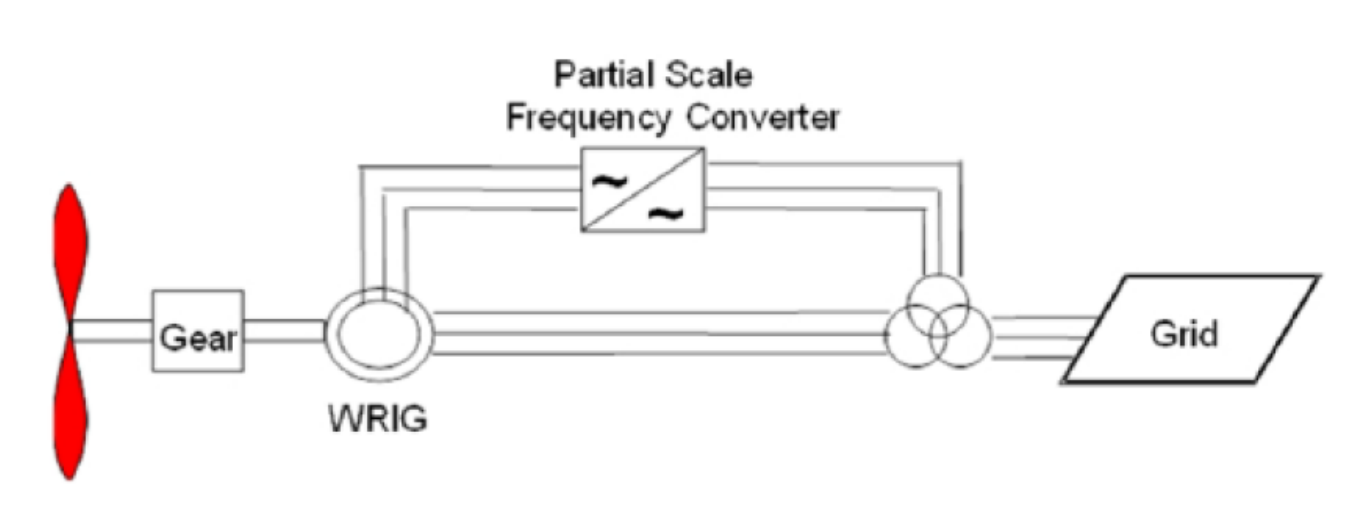}%
\label{fig_first_case}} 
\subfloat[Type IV]{\includegraphics[scale=0.35]{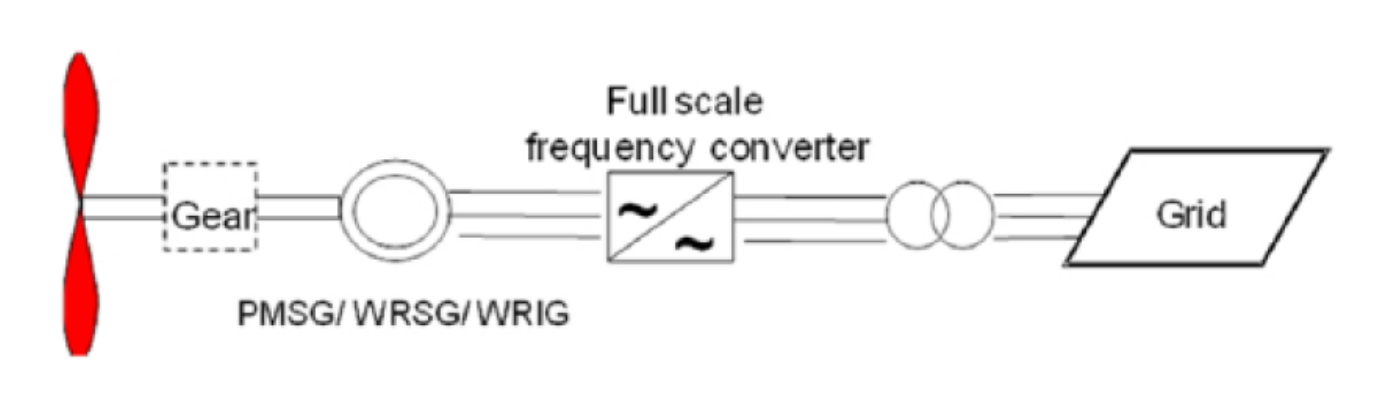}%
\label{fig_second_case}}
\caption{IEC classification of WTs \cite{DTUwind}.}
\label{WTtypes}
\end{figure}

Fig. \ref{WTtypes} displays the various types of wind turbines (WT) based on the IEC-61400-27-1 standard. The detailed dynamic models can be found in reference \cite{WTdynamics}.

Type I and II are fixed speed operation, while type III and IV are variable speed operation which represent a great advantage considering the dominance of variable speed operation of the WTs. Nevertheless, each type has important pros and cons listed below \cite{DTUwind, WTenergies, WTdynamics, WTcomparative}:

\begin{itemize}
    \item Design and capacity: Type III are typically designed for medium wind speeds with capacities ranging from a few hundred of kW to MW, while type IV are designed for higher speeds and larger capacities.
    \item Grid connection: Type IV often use variable speed generators and may incorporate power electronics for grid connection that allows to optimize the output power. 
    \item Rotor design: Type III has a simpler rotor design compared with type IV, which often feature more advanced aerodynamic features and control mechanisms.
    \item Grid support features: Type IV turbines may include advanced grid support features such as low-voltage ride-through capability and active power control, enhancing the stability of the grid.
\end{itemize}

\subsubsection*{Hydroelectric Power Systems}

Hydroelectric power plants harness the energy of flowing water to generate electricity. Power electronic converters could be used in variable-speed hydro turbines and pumped-storage hydroelectric plants to adjust the speed and frequency of generator output to match grid requirements \cite{hydro}.

\subsubsection*{Ocean Energy Systems}

Ocean energy technologies, such as tidal turbines, wave energy converters, and ocean thermal energy conversion (OTEC) systems, utilize power electronic converters to convert the variable mechanical energy of ocean waves, tides, and temperature differences into grid-compatible electricity \cite{ocean}.

\sectionbreak
\section{IED Protection Challenges}
\label{sec:IED_challenges}

The protection system is a crucial element of the power system and  has the objective of protecting people, equipment and the power system integrity. 

Relays or Intelligent Electronic Devices (IED) are core to the  protection system. It is designed to detect faults and send the appropriate signals to circuit breakers to isolate the faulted parts of the grid. Doing so, severe damage to primary equipment is prevented and the healthy part of power system is separated from the faulted part, avoiding black-outs. In order to successfully clear the fault, the relay protection must meet several criteria.

The main criteria for relay protection include:
\begin{description}
\item[Selectivity:] The protective relay should be capable to define whether the fault occurs within the protected zone or beyond it and operate in a correct order and time to coordinate with other relays.

\item[Sensitivity:] Relays are able to detect and respond to changes in electrical parameters that indicate the presence of a fault or abnormal condition in the power system.

\item[Speed:] Relay protection device must be able to detect the fault and send tripping signal to circuit breaker in a specific range of time.
    
\end{description}

Modern numerical relay protection devices are based on microprocessor technology and will be referred as intelligent electronic devices (IEDs) in this report.

The specific features of converter-dominated system mentioned in chapter 2 can make a great impact on the IEDs in the power system. This fact may impose challenges on reliable operation of IEDs during fault events.

The Table \ref{table_IED_challenges} represents the overview of main issues raised by different authors in the last 10 years.

\begin{small}
\begin{table}[H]
\caption{IED Protection Challenges}
\label{table_IED_challenges}
\centering
\begin{tabular}{lccccccccccc}
\hline
\textbf{Challenge} & \textbf{\cite{DirectIED}} & \textbf{\cite{NegSeqIED}} & \textbf{\cite{SwingIED}} & \textbf{\cite{CurrentIED}} & \textbf{\cite{CurImpIED}} & \textbf{\cite{CurImpIED2}} & \textbf{\cite{OverviewIED}} & \textbf{\cite{PhaseIED}} & \textbf{\cite{SwingIED2}} & \textbf{\cite{rocofIED}} & \textbf{\cite{entsoeIED}} \\ 
\hline \hline
\multicolumn{1}{c}{\textbf{Distance}}  & &  &  &  &  &  &  &  &  &  & \\
 Starting current        &  $\times$ & $\times$ & $\times$ & $\times$ & $\checkmark$ & $\checkmark$ & $\checkmark$ & $\checkmark$ & $\times$ & $\times$ & $\times$ \\
 Impedance reach         &  $\times$ & $\times$ & $\times$ & $\checkmark$ & $\checkmark$ & $\checkmark$ & $\checkmark$ & $\checkmark$ & $\times$ & $\times$ & $\times$\\
Directionality          &  $\checkmark$ & $\times$ & $\times$ & $\times$ & $\times$ & $\times$ & $\checkmark$ & $\times$ & $\times$ & $\times$ & $\times$\\               
 Phase selection         &  $\times$ & $\checkmark$ & $\times$ & $\times$ & $\times$ & $\times$ & $\checkmark$ & $\times$ & $\times$ & $\times$ & $\times$\\
 \hline 
 \multicolumn{1}{c}{\textbf{Negative sequence}} & &  &  &  &  &  &  &  &  &  & \\
Overcurrent             &  $\times$ & $\checkmark$ & $\times$ & $\times$ & $\times$ & $\times$ & $\checkmark$ & $\checkmark$ & $\times$ & $\checkmark$ & $\times$\\
Directional overcurrent &  $\times$ & $\checkmark$ & $\times$ & $\times$ & $\times$ & $\times$ & $\checkmark$ & $\checkmark$ & $\times$ & $\checkmark$ & $\times$\\
Starting logic element  &  $\times$ & $\times$ & $\times$ & $\times$ & $\times$ & $\times$ & $\checkmark$ & $\times$ & $\times$ & $\times$ & $\times$\\
 \hline 
  \multicolumn{1}{c}{\textbf{RoCoF}} & &  &  &  &  &  &  &  &  &  & \\
Power swing blocking    &  $\times$ & $\times$ & $\checkmark$ & $\times$ & $\times$ & $\times$ & $\checkmark$ & $\checkmark$ & $\checkmark$ & $\checkmark$ & $\times$\\
Load shedding           &  $\times$ & $\times$ & $\times$ & $\times$ & $\times$ & $\times$ & $\times$ & $\times$ & $\times$ & $\times$ & $\checkmark$\\
Out-of-step             &  $\times$ & $\times$ & $\checkmark$ & $\times$ & $\times$ & $\times$ & $\checkmark$ & $\checkmark$ & $\checkmark$ & $\checkmark$ & $\times$\\
\hline
\end{tabular}
\end{table}
\end{small}

\subsection{Distance protection}
\subsubsection{Principles}
\begin{figure}[H]
        \centering
        \includegraphics[scale=1]{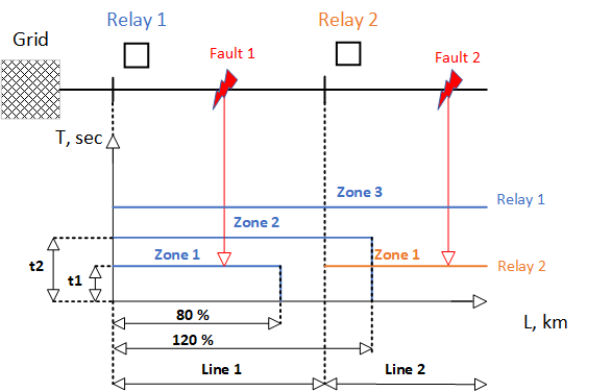}
        \caption{Distance protection zones} 
        \label{figurezones}
    \end{figure}

Distance protection is commonly used as backup or main protection  for the fault identification in high voltage transmission lines. Distance relay operation is based on calculation of the impedance using the measurements of line voltage and current. Distance protection evaluates the measured impedance (measured voltage divided by measured current). In case of a short circuit, the line impedance will dominate and the measured impedance will be very inductive and low. While in case of normal operation, the impedance will be higher and more resistive due to the load.  When a fault occurs, the impedance seen by the relay will shift, which can be detected by the relay. The first zone usually protects 80\% of the line also known as under-reaching, the second zone (overreaching) used to cover 120\% of the line and create a local back up for next segment of the grid as shown in the Figure \ref{figurezones}.

In recent years, power systems with a high share of converter-based energy sources have unveiled numerous challenges for distance protection function. These challenges are related to issues with overcurrent starting, impedance reach accuracy, directional fault discrimination and phase selection error.

\subsubsection{Overcurrent starting}

Starting function or fault detection is used to detect and classify faults in the grid. Pick-up and drop-off actions of the starting function are marking out the start and the finish of the fault event. It also activates the set of timers, other protection functions and the fault recorder.

One of the approaches for starting functions is over-current starting.
The starting current of a distance relay is set to the minimum fault current level required to initiate the operation of the relay.

The fault current provided by grid-side converters is limited to 1.1-1.2 p.u. by converter controls to avoid damage of the power electronic switches. Such current level is very close to the nominal load current. In case of converter-dominated power system the low current infeed from converters can compromise the starting unit of relay. \cite{CurImpIED, CurImpIED2, OverviewIED, PhaseIED}

\subsubsection{Impedance reach accuracy}

A distance relay is configured to operate within a specified impedance range. The relay remains inactive while  calculated impedance exceeds the predetermined value. Such threshold value of impedance (or the corresponding distance) is referred to as impedance reach of distance relay. 

\begin{figure}[H]
        \centering
        \includegraphics[scale=0.5]{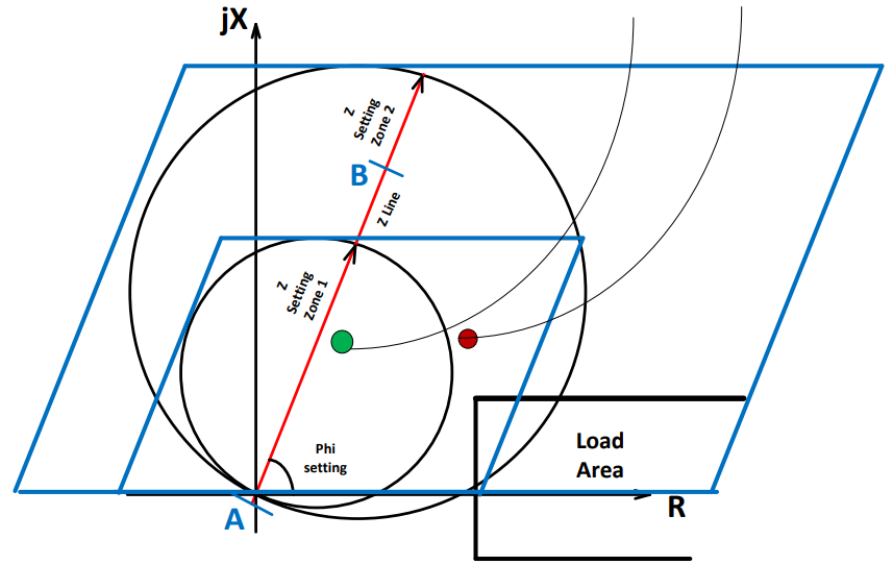}
        \caption{Impedance reach accuracy} 
        \label{figureimpcharact}
    \end{figure}

The distance protection placed close to the converter-based sources may have a big deviation of the calculated impedance value and relay may not fulfill the criterion of selectivity and trip the zone 2 instead of zone 1 as shown in Figure \ref{figureimpcharact}.

The malfunction of protection devices due to this issue was reported in \cite{CurrentIED,CurImpIED, CurImpIED2, OverviewIED, PhaseIED}

\subsubsection{Directional fault discrimination}

Directional fault discrimination in distance relays refers to its ability to determine the direction from which a fault or disturbance originates in the power system. This functionality enables the relay to selectively trip and operate only for faults occurring in a specific direction, ensuring selectivity criterion of protection.

There are several main approaches to numerical direction determination also referred as polarisation:

\begin{itemize}
\item Self-polarisation.
\item Cross-polarisation.
\item Positive/Zero sequence polarisation.
\end{itemize}

Cross-polarization and positive sequence polarization direction fault discrimination algorithms may be compromised during single-line-to-ground faults in the offshore wind farms as was reported in \cite{DirectIED} due to various system parameters of offshore wind farm collection system.

\subsubsection{Phase selection}

Phase selection in distance protection involves the process of identifying the fault type: phase-to-ground, two-phase-to-ground, phase-to-phase or three-phase fault and the faulted phase. It is implemented in IED by comparing sequence currents during the fault in the system: positive, negative and zero sequence.

Since the current output from converter-based source is fully controlled by converter controls and its behavior must comply  with fault-ride-through requirement in Grid Codes the wrong fault identification may appear \cite{NegSeqIED, OverviewIED}. The main reason of it is current limitations of converter switches and suppression of negative sequence currents by converter controls.

\subsection{Negative sequence based protection}

Negative sequence-based protection relies on the detection of a negative sequence component of  voltage or current during (unbalanced) faults. In a balanced system, the negative sequence component is typically close to zero. However, during unbalanced conditions, such as phase-to-phase faults or phase-to-ground faults, the negative sequence component becomes significant.

By tracking the negative sequence component, IEDs can detect unbalanced conditions and isolate the affected section of the power system. Negative sequence-based protection is commonly used as a back-up protection in power systems or as a starting logic element in other relay protection functions.

\subsubsection{Overcurrent}

Negative sequence overcurrent protection operates by comparing the magnitude of the negative sequence current to a predetermined threshold. If the negative sequence current exceeds this threshold, indicating the presence of an unbalanced fault, the protection function initiates a tripping signal.

In case of converter-dominated system, grid-side control of converter-based source may suppress the negative sequence current to zero. It depends on the protection philosophies and fault-ride-through requirements of specific country. In this scenario the overcurrent negative sequence protection operation may be compromised as shown in \cite{NegSeqIED, OverviewIED} due to the lack of negative sequence current injected to the grid.

\subsubsection{Directional overcurrent}

Unlike conventional overcurrent protection, which responds solely based on the current magnitude, directional overcurrent protection operates by discriminating between fault currents flowing into or out of a protected zone.
In case of system topology where current infeed to protected grid is possible from several sides the directional protection function is commonly used as a back-up protection.

The directional characteristic of a negative sequence overcurrent relay is defined by two key factors: the operating quantity, typically the input negative sequence current, and the polarizing quantity. The polarizing quantity is commonly a negative sequence voltage obtained by rotating the input voltage by the maximum torque angle. The directional algorithm computes the angle between the polarizing quantity and the operating quantity. Initially, the polarizing quantity is rotated by the maximum torque angle. Subsequently, if the angle is smaller than the angle defined by the operating sector value, the current is identified as forward current. 

References \cite{NegSeqIED, OverviewIED} have explored the possible test scenarios leading to maloperation of directional negative sequence overcurrent relays in the grid with converter-based sources. This type of protection function may fail because its operation principle is based on the angle between negative current and voltage values, which is deviating a lot due to grid-side converter controls.

\subsubsection{Starting logic element}

One of the applications for negative sequence based protection is a starting function in communication-based protection schemes. As a unit inside of a large chain of logic elements the failure to pick-up the fault can lead to wrong operation of the whole protection function in IED\cite{OverviewIED}.

\subsection{ROCOF based protection}
The primary objective of ROCOF based protection is to detect and mitigate the effects of disturbances such as generator loss, load shedding, or sudden changes in demand that can cause rapid frequency fluctuations in the power system. These disturbances can potentially destabilize the grid and lead to widespread blackouts if not promptly addressed.

\subsubsection{Power swing blocking}

The power swing mode is a power system mode in which power system parameters are changing periodically without breaking synchronisation of the system. Power swings can lead to fluctuations in voltage and current levels, and may pose challenges to the stability and reliability of the power system. It's important for relay protection device, such as distance relay, to differentiate between power swings and actual faults to avoid unnecessary tripping of circuit breakers and maintain the integrity of power system.

    \begin{figure}[H]
        \centering
        \includegraphics[scale=0.45]{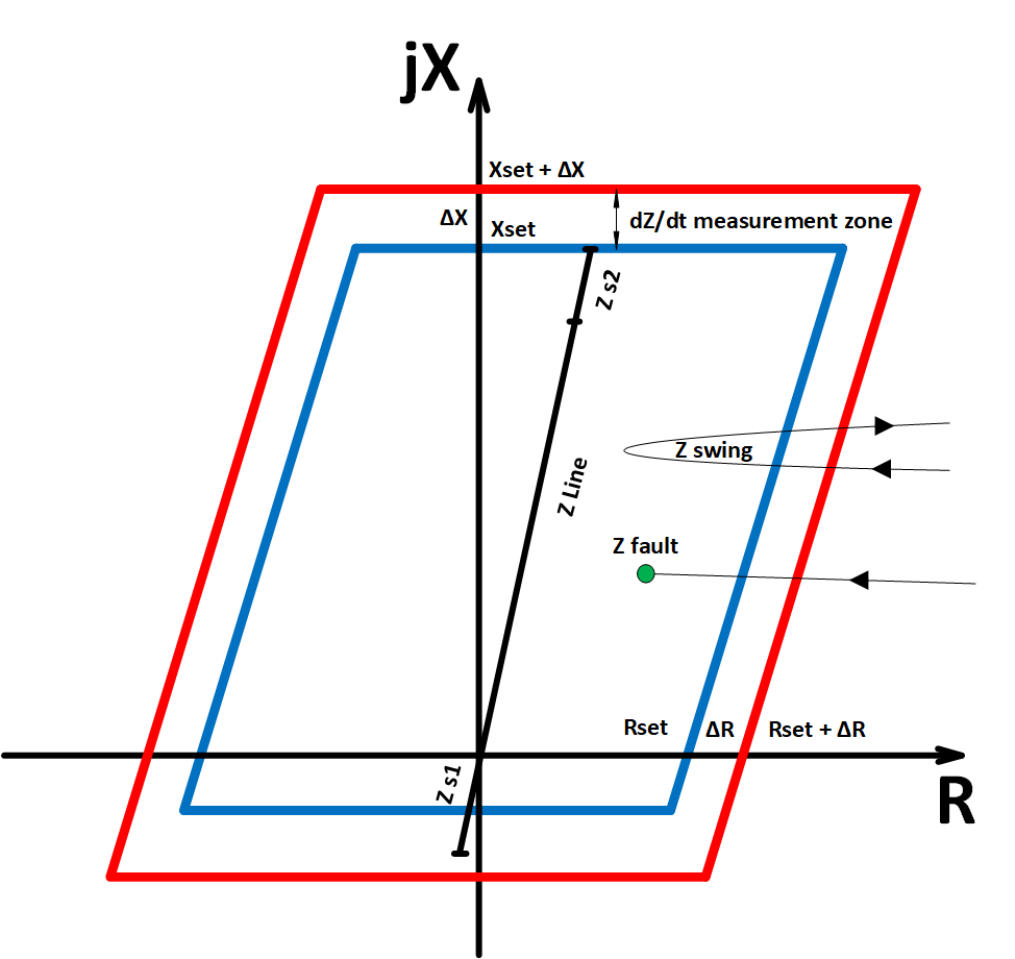}
        \caption{Power swing blocking} 
        \label{figurepowerswing}
    \end{figure}

In order to discriminate faults and power swings, a power swing blocking function is implemented in distance relays. There are several approaches commonly used by relay manufactures, e.g. rate of change of positive sequence current, rate of change of negative sequence current etc. One of the most popular logic behind power swing blocking function relies on the rate-of-change-of-frequency (ROCOF).

During the fault, measured impedance drastically decreases and jumps inside distance protection operation zone with high ROCOF, while during power swing mode the impedance trajectory is gradually falling into the protection zone at frequency of system power swing. dZ/dt value is calculated by measuring the time required for the impedance vector to cross an area between two impedance zones as shown in figure \ref{figurepowerswing}.

With increasing share of converter-based sources in the system, the ROCOF value is increasing and power swing blocking function may maloperate as reported in \cite{SwingIED, OverviewIED, PhaseIED, SwingIED2, rocofIED}. It becomes harder for IED to distinct ROCOF caused by a power swing from that of a fault.

\subsubsection{Out-of-step protection}

\begin{figure}[H]
        \centering
        \includegraphics[scale=0.45]{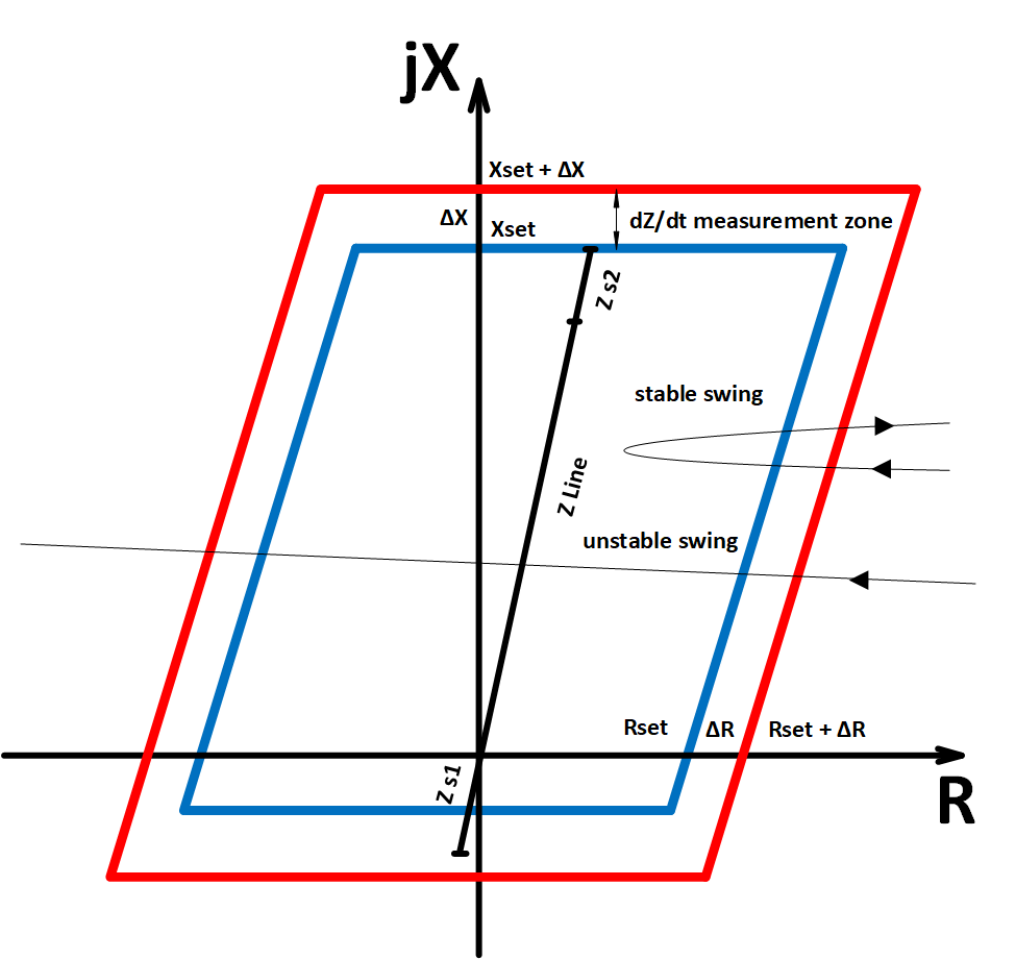}
        \caption{Out-of-step protection} 
        \label{figureunstablepowerswing}
    \end{figure}

Out-of-step protection is using a similar principle as power swing blocking function to distinct between stable and unstable power swings as represented in Figure \ref{figureunstablepowerswing}.

References \cite{SwingIED, OverviewIED, PhaseIED, SwingIED2, rocofIED} reported on possible scenarios of out-of-step protection maloperation as a consequence of high ROCOF in the system with high penetration of converter-based sources.

\subsubsection{Load shedding}

Under-frequency load shedding (UFLS) is a relay protection function implemented in electrical power systems to maintain system stability and prevent widespread blackouts and frequency collapse during periods of insufficient power generation. UFLS is typically implemented in a hierarchical manner, with multiple levels or stages of load shedding triggered at different frequency thresholds. 

Some of UFLS logic is based on an additional ROCOF measurement. Reduced inertia and faster transient dynamics may lead to inaccuracy of ROCOF measurements and unintentional tripping of UFLS in converter-dominated system. \cite{entsoeIED}

\sectionbreak
\section{Converter Challenges}
\label{sec:converter_challenges}

The advancement in converter technology as well as continuous decrease in cost of energy from renewable energy sources like solar PV and wind  has made a converter dominated power system a possibility. But it is still very challenging to achieve a converter dominated power system because of issues caused by different behaviour of the converters and converter dominated system as compared to the conventional systems with SGs. This results in a more complex converter control as the transition to a converter dominated system progresses. In this section, the protection challenges due to the behaviour and control mechanisms of the converters is discussed. Some of the recent issues have been categorized in Table \ref{converter_control_challenges}.

\begin{table} [h!]
\begin{small}
\caption{Converter control challenges posed by 100\% converter-based systems}
\label{converter_control_challenges}
\centering
\begin{tabular}{lcccccccccc}
\hline
\textbf{Subject}  & \textbf{\cite{43}} & \textbf{\cite{44}} & \textbf{\cite{45}} & \textbf{\cite{46}} & \textbf{\cite{47}} & \textbf{\cite{49}} & \textbf{\cite{50}} & \textbf{\cite{51}} & \textbf{\cite{52}}\\ 
\hline
Dynamic voltage support & $\checkmark$ 
& $\checkmark$ & $\checkmark$ & $\times$ & $\times$ & $\times$ & $\times$ & $\times$ & $\times$ \\
\hline
\multicolumn{1}{c}{\textbf{Stability \& control}} &  & & &  &  &  &  &  &  \\ 
Loss of synchronization & $\times$ & $\times$
& $\times$  & $\checkmark$ & $\times$ & $\times$ & $\times$ & $\times$ & $\times$ \\
Impact of current control & $\times$ & $\times$
& $\times$ & $\times$  & $\checkmark$ & $\checkmark$ & $\checkmark$ & $\times$ & $\times$\\
Control parameters' influence & $\times$ & $\times$ & $\times$ & $\times$ & $\times$ & $\times$ & $\times$ & $\checkmark$ & $\checkmark$ \\
\hline
\end{tabular}
\end{small}
\end{table}

\subsection{Dynamic Voltage Support capability of Converters}

When a fault occurs in a grid network, it leads to a reduction in voltage also known as a `voltage sag'`or `voltage dip'. Grid faults can be divided into two types: Symmetrical faults (three phase and three phase to ground faults) which are the most detrimental and unsymmetrical faults(single line to ground and line to line fault)  which are more likely to occur. To maintain the grid voltage during such times of voltage sag and unbalanced voltages, reactive current, both in positive and negative sequence need to be injected. This is known as dynamic voltage support.

In conventional systems with SGs, a strong fault current is injected into the system in order to keep the voltage at acceptable levels. The voltage is measured at the SG terminals and compared against a reference value. Then, the AVR can adjust the excitation level to generate the required reactive current needed to keep the voltage within the specified limits. But, it is a very different situation in converter dominated system. Unlike SGs, converters have a very limited short circuit contribution of 1-1.2pu as their control uses a current limitation algorithm that keeps the converter current near to the rated value in order to protect them. This current limitation limits the reactive current injection by converters and in converter dominated systems and the grid can suffer from more severe voltage sags to a level that undervoltage protection could be activated and network devices are interrupted because of large voltage deviation.

The converter behaviour mentioned above poses a great challenge to the protection system. The widely used overcurrent protection may not work properly due to the current limitation required by the converters. The undervoltage protection provides an alternative to the overcurrent protection as it is independent of the magnitude and direction of the fault current. It functions in a way wherein it is activated when the voltage falls below a certain threshold. But, as mentioned in \cite{44},  it is difficult to detect whether the voltage drop is due to fault or sudden change in loads or generation loss. For a fault at the end of a long feeder, the voltage does not dip significantly at PCC. Under this situation, the undervoltage protection scheme cannot provide enough selectivity and hence, is unreliable.

Some of these issues were tackled by introducing new grid code requirements for LVRT as described in \ref{sec:gridcode} and reactive current injection during faults. The UK grid code requires that the reactive current in both positive and negative sequence be evenly reduced in case of current saturation \cite{43}. Fast fault current injection (FFCI) is another vital requirement in the grid codes, which facilitates protection systems to detect fault and support system voltage during the faults and system voltage recovery following the fault clearance. Existing grid code requirements
on FFCI vary significantly across European countries. In these grid codes, a dead zone setting is typically adopted, which is in the range of 0.9-1.1 p.u. of the Point of Common Coupling (PCC) voltage. When the system voltage drops below 0.9 p.u., power generation modules are typically required to inject reactive current proportional to the level of voltage depression. With respect to the injection proportion, for example, UK and Denmark require 2.5\% reactive current injection for 1\% PCC voltage depression, while German grid code requires 2\% reactive current injection for 1\% PCC voltage reduction \cite{45}.

\subsection {Stability and control}

Transient stability refers to the ability of the system to operate in synchronization following a large disturbance (a fault, a sudden change in load, a sudden generation loss).

Converter dominated systems are characterized by low inertia and decreased damping. This leads to issues that arises with higher ROCOF and a lower frequency nadir as mentioned in section \ref{sec:special_aspects}. Moreover, the dynamics of the converters are much different and especially much faster than conventional SGs. This presents a challenge for the converter control and for the transient stability of the system, but also an opportunity as fast controls allow to improve stability.

Some of the challenges posed by the complexity of converter control to transient stability are provided below.

\subsubsection{Loss of synchronization}

While the synchronization in conventional power system with SGs is provided by the inherent synchronization mechanism of synchronous generator, the synchronization of converters is accomplished by detecting and following the phase angle of the terminal voltage. This phase angle  is generally obtained through a phase locked loop (PLL) or a power synchronization loop depending on the control. 

Synchronization is a major issue in converter dominated systems, especially for the IBRs connected to weak grids as the voltage magnitude and phase at the PCC is much more sensitive to the variation of grid operating points. The loss of synchronization leads to a current injection at a frequency different from the grid fundamental frequency by the IBRs. This in turn leads to overcurrents which can force the IBRs to disconnect and ultimately lead to  cascaded outages and even a blackout of the converter dominated system. Normally, the grid code requires converters to remain connected and support the grid. One of these requirements is the LVRT where the converter has to remain connected to the grid after the fault even when the grid voltage is significantly reduced at the device terminals up to a given limit in both magnitude and time.

As mentioned earlier, there are two ways converters are synchronized to the grid. With PLL and without PLL. GFLs, which are widely used are normally synchronized through a PLL while grid-forming converters (GFM), which are becoming more and more popular, can be synchronized with or without a PLL (see also section \ref{sec:solutions}). GFMs have been shown to work in stand alone mode as well as grid connected mode. So, once the output voltage of the GFM converter is synchronized with the grid voltage it is expected to seamlessly transition between the two modes. But, as explained in \cite{46} this is not the case as  the GFM inverter may still lose its synchronization with the main grid after the re-connection with grid, and such loss of synchronization is attributed to the transient instability of GFM inverters. It has been shown that during the transition from islanded mode to grid connected mode, the loss of synchronization occurs due to the bypass of the resynchonization power controller.

\subsubsection{Impact of current control}

Compared to SG that can support up to seven times over its rated current, power converters can only cope with limited amount of overcurrent which is 1-1.2 times of the rated value. Trying to mimic synchronous machine behaviour would require a very large oversizing of semiconductor components and induce large additional costs \cite{50}. Therefore, power converters have to be protected against overcurrents due to large disturbances like line tripping and connection/disconnection of a large load. Different methods for current limitation have been proposed in the literature for current limitation in VSC. The two most widely adopted methods in literature are based on current saturation algorithm (CSA) or current reference limiting and virtual impedance (VI) \cite{50}.

In \cite{47}, a current reference limiting strategy has been used to study the transient stability of a P-f droop controlled VSC. It has been shown that the current limitation has a direct impact on the critical clearing angle. Critical clearing angle is the angle in the $P-\delta$ curve beyond which the system goes into instability. In the same article it has been shown that the margin for critical clearing angle decreases due to the current limitation during large disturbances and the VSC can go into transient instability. 

In \cite{49}, a similar study with droop controlled VSC has been done with virtual impedance (VI) based current limitation. It has been shown with the analysis of critical clearing time during a large disturbance that the VI based current limitation can improve the transient stability of the system. In \cite{50}, the transient stability of a system  has been compared with CSA and VI based current limitation and it was concluded that while CSA provides better current limitation the VI provides a better transient stability.

\begin{figure}[H]
        \centering
        \includegraphics[scale=0.25]{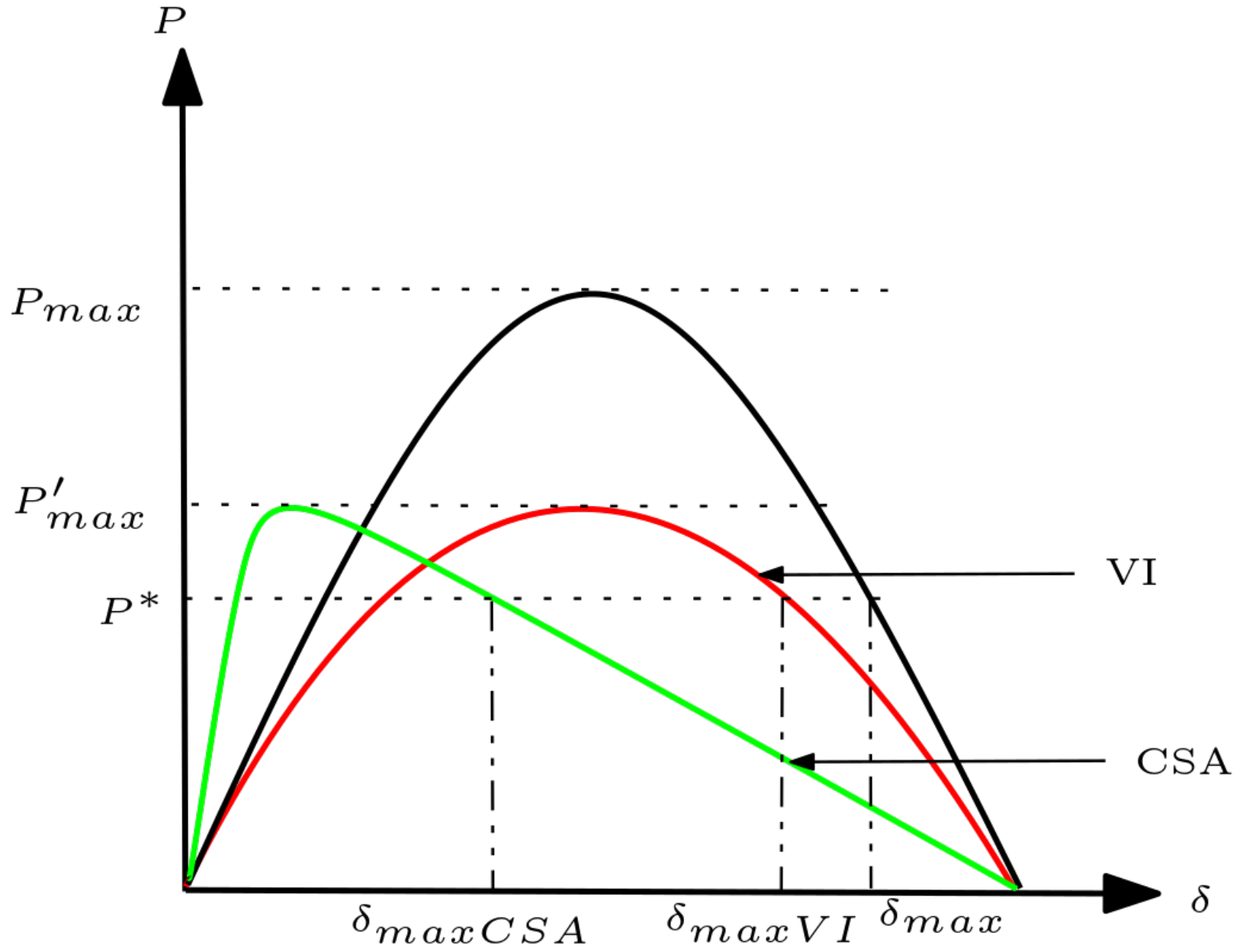}
        \caption{P-$delta$ curve for different current limitation techniques} 
        \label{transient stability}
\end{figure}

It can be summarized that although different current limitation techniques can provide varying level of transient stability, they certainly have an impact on the critical clearing angle/time, which are a deterministic factor in the design of protection system. Thus, special attention needs to be paid while designing the protection system considering the transient stability limits for converter based systems.

\subsubsection{Influence of control parameters on transient stability}

It has been summarized in section\ref{sec:special_aspects} how the converter dominated systems differs from the conventional system characterized by SGs because of the different inherent characteristics of the converter. Unlike SGs, converters do not have the inherent capabilities to provide inertia, damping and short circuit capability. But the same limitations also provide higher degrees of freedom to implement control functions in order to achieve the same service provisions as SGs. The converters are accompanied with different control systems such as current limitation, active power loop etc in order to function like SGs and at times provide even better performance than SGs. 

The inertia or virtual inertia and damping parameters of the converters are not inherent characteristics and are rather controllable parameters. Thus, they provide flexibility of operation since the control variables can be tuned according to the requirement. Since these control variables and control systems are used to replicate the characteristics provided by SGs, they will undoubtedly influence the dynamics and transient stability of the system.

In \cite{51}, it has been shown how the parameters of the active power loop of the grid forming converter effects the transient stability of the systems. The focus has been in the damping parameter, which is kept much higher in converters than the damping parameter of SGs, and the virtual inertia. They have concluded that both the damping parameter and virtual inertia impact the critical clearing angle (CCA) and critical clearing time (CCT) during when a fault occurs. The study showed that greater the damping parameters the greater the CCA and CCT, and the greater the virtual inertia the smaller the CCA and the greater the CCT. Moreover, further study showed that for a smaller converter active power capacity with a large damping parameter and under lighter faults, there may be abnormal behaviour in the effect of damping parameter on CCA and virtual inertia on CCT, i.e., as the damping parameter increases, CCA decreases, and as virtual inertia increases, CCT decreases.

A similar study is done in \cite{52}, where the transient stability of a droop controlled VSC is analyzed through the influence of both $P-f$ and $Q-V$ droop parameters on the CCA and CCT, with a special focus on the effect of the $Q-V$ droop. It has concluded that the transient stability of the VSC is better with higher values of both $P-f$ and $Q-V$ droop coefficients.
\sectionbreak
\section{Grid Code requirements overview}
\label{sec:gridcode}
\begin{figure}[H]
        \centering
        \includegraphics[scale=0.4]{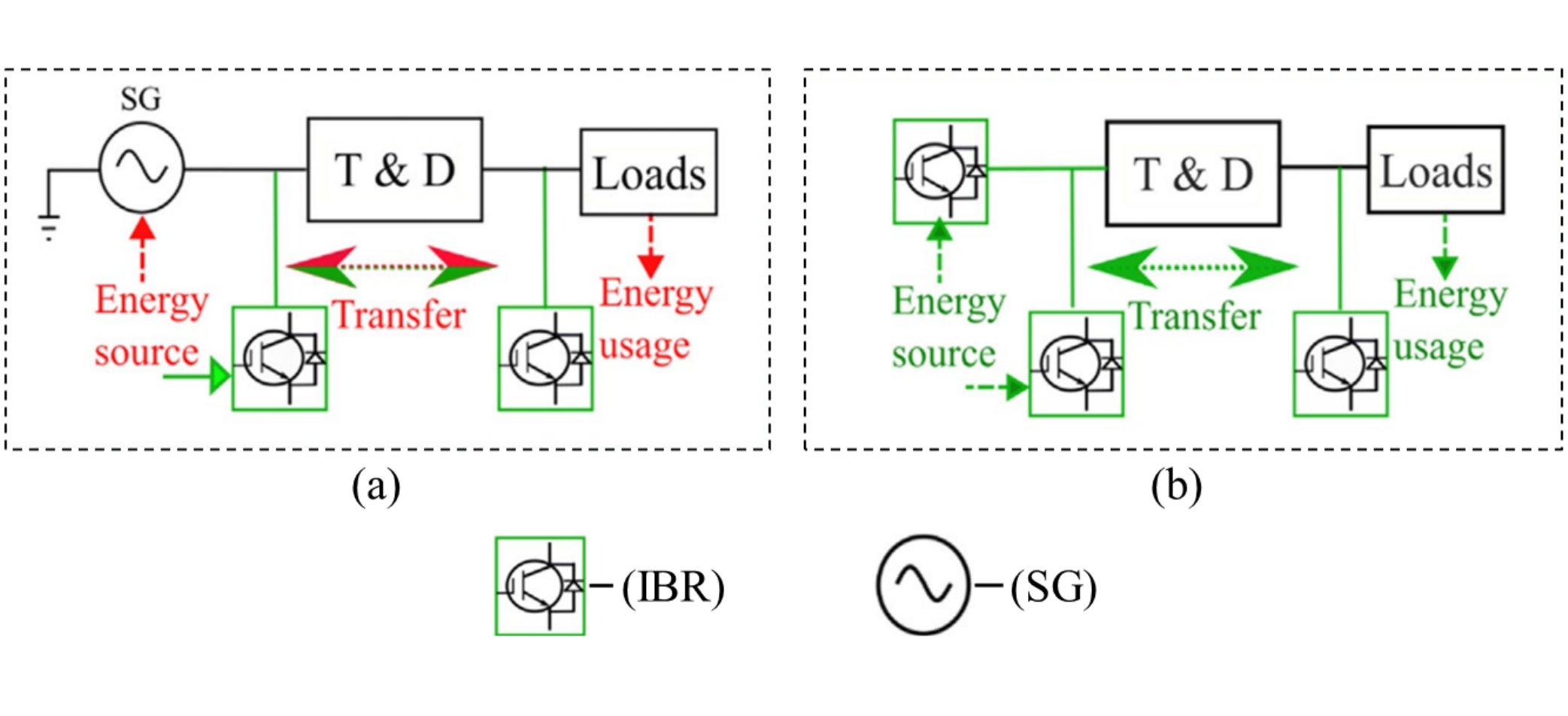}
        \caption{Power system (a) at present, (b) in future with 100\% IBR} 
        \label{Psys}
    \end{figure}

The electric power system is a complex interconnection of generators and loads, connected via transmission and distribution lines. It plays a crucial role in safety, economy, and national development. Grid codes, which are rules and regulations governing the operation of the grid, have evolved over time to accommodate technological advancements and changing energy sources. Challenges arise from integrating renewable energy into existing grids, required collaborative efforts among stakeholders \cite{international2021assessing}. In a grid with increased renewable energy penetration, handling the intermittent nature of renewables becomes crucial. Therefore, accurate prediction and effective control are necessary to reduce their impact on the grid’s stability. Modern grid codes objectives are not only to address the conventional generators and loads but also to support the integration of renewables With the aim of achieving 100\% renewable energy, the future power system will transform to inverter-based resources as shown in Figure \ref{Psys} \cite{sathiyanarayanan2024grid}.

\subsection{Development and technical requirements of grid code}

Grid code development involves policy mandates, stakeholder consultation, technical studies, authorities approval and regular revision to adapt changing system needs and technological advancements. The current grid codes, which address the issues such as fault ride-through, fault location, and system restoration, demonstrate how grid code standards have evolved over the time. Grid codes are not universal and differ between nations and regions based on their operations, requirements, and grid characteristics. Additionally, they change throughout time to take into account new advancements in the power industry and technologies. Therefore, establishing the grid code is crucial as it regulates the grid access regardless of whether the power system is operated by a specific entity or a vertically integrated utility \cite{tsili2009review}.

\begin{figure}[H]
        \centering
        \includegraphics[scale=0.4]{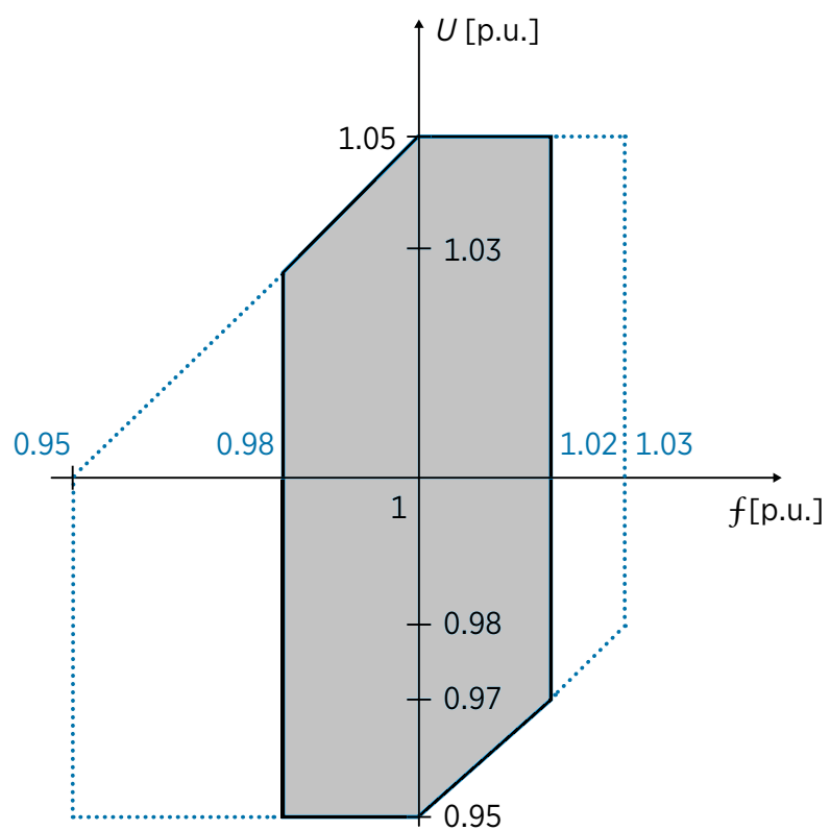}
        \caption{IEEE C50.13 and IEC60034-3 limits on voltage and frequency} 
        \label{VFlimits}
    \end{figure}

\begin{figure}[H]
        \centering
        \includegraphics[scale=0.4]{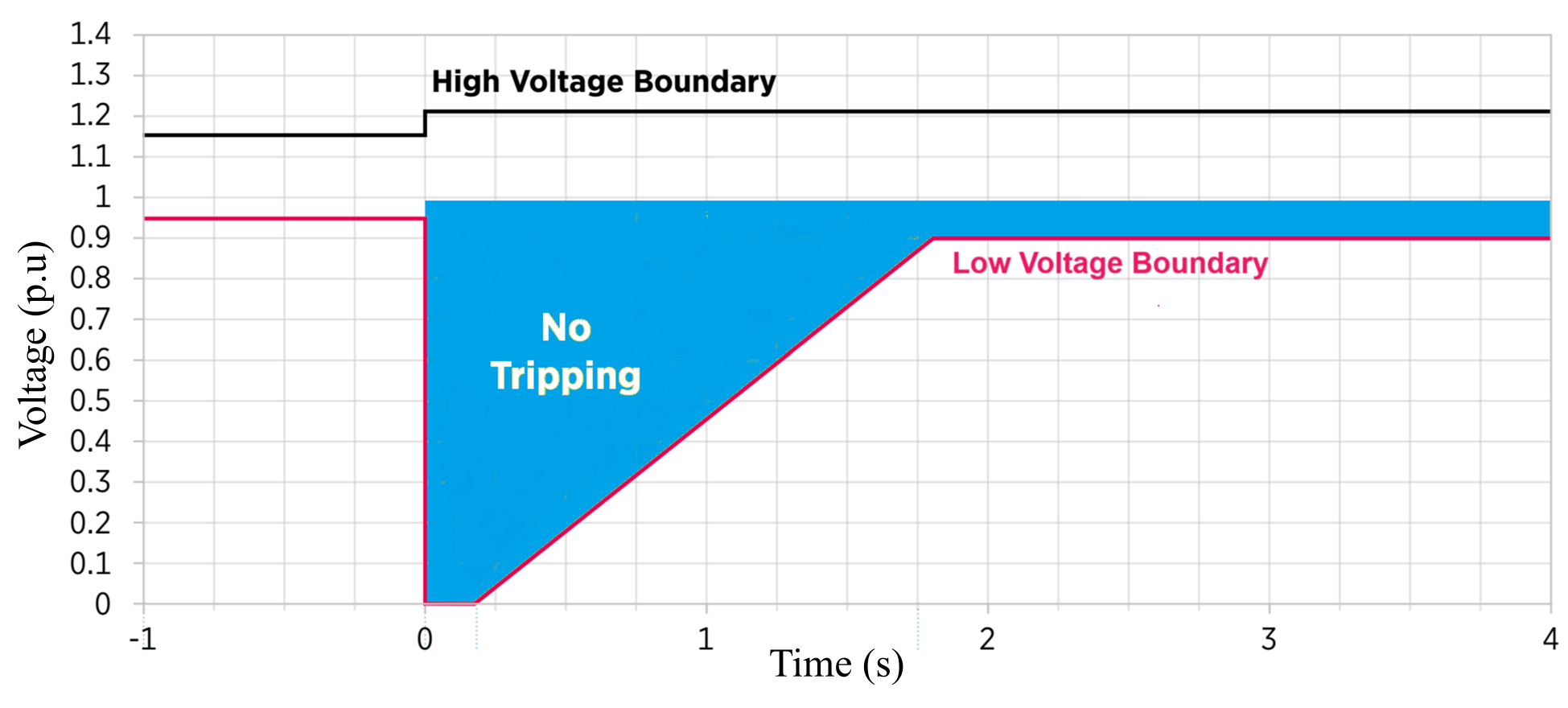}
        \caption{Voltage ride through requirements} 
        \label{VRT}
    \end{figure}

The technical requirements of grid codes specify the functional variables of the system that the renewable energy sources (RES) must comply with, such as voltage, frequency, active and reactive power, power quality, fault ride-through, etc. These requirements vary according to different regions and countries, and are periodically updated to reflect the changing grid conditions and technologies. The integration of RES must undergo various tests and certification processes to ensure their grid compliance and performance. Voltage and frequency support is essential for the stability of the power system \cite{khalid2022efficient}. RES must provide voltage and frequency support during normal and abnormal conditions, such as faults, load changes, or generation variations. This includes high voltage ride through (HVRT) and low voltage ride through (LVRT) capabilities (illustrated in Figure \ref{VRT}) , as well as inertial response and frequency control \cite{erlich2005grid}. The grid codes define the specific values and limits for the voltage and frequency deviations and recovery times for different scenarios. Generally, generators must be able to operate continuously in specific voltage and frequency ranges as shown in Figure \ref{VFlimits} (for instance, 0.95 to 1.05 per unit voltage and ± 1 Hz in 50 Hz frequency), as well as maintain a connection within a defined range.

Active and reactive power control are needed to regulate the power flow and voltage in the system, as well as to coordinate the operation and dispatch of multiple RES. RES must be able to follow the set points and commands from the grid authority or operator, and adjust their output according to the system needs and market signals. RES must also be able to provide ancillary services, such as voltage regulation, frequency regulation, spinning reserve, etc. Fault ride-through and system restoration are the abilities of RES to withstand and recover from faults or disturbances in the system, and to support the system recovery and restoration. RES must not disconnect from the grid during faults, unless specified by the grid codes, and must help to clear the faults and restore the system voltage and frequency. RES must also have black start or grid-forming capabilities, which enable them to start or form a grid without external power sources. 

\begin{table}[H]
\caption{LVRT requirements for different countries \cite{irena2022grid}}
\label{LVRT}
\centering
\begin{tabular}{|l|l|l|l|l|l|l|l|l|}
\hline
Parameter & UK & Denmark & Ireland & Egypt & France & Italy & China & Germany \\
\hline
\(t=t_r\) (ms)  & 140  & 100   & 625    & 250   & 200    & 500   & 625   & 150     \\
\(V=V_r\) (p.u.)&0.15&0.20&0.15&0&0.20&0.20&0.20&0\\
\(t=t_s\)       &500    &600 &3000 &3000 &2000 &300 &2000 &1500\\
\(V=V_s\) (p.u.)&0.94    &0.75 &0.90     &0.90 &0.85 &0.75 &0.90 &0.90\\
\hline
\end{tabular}\\
\end{table}

\subsection{Grid codes for 100\% RES}

The rise of renewable energy sources is reshaping the power grid, shifting from its dependence on synchronous machines to an inverter/converter-dominated electric grid. However, the transition to a fully renewable-based system will face significant challenges because of our dependence on rotational machines, already well-established protection for synchronous machines, and certainty provided by conventional source-based systems. To address these challenges, strategies including optimised control of inverter-based systems to emulate synchronous machine behaviour, utilising energy storage systems to support RES and enhance dispatchability, alongside emerging technologies like the Internet of Things, to effectively operate within a 100\% renewable-based system with comparable or improved performance.

Modern grid codes must incorporate various functions based on control logic, converter topology, energy storage characteristics, and protective measures. Unlike conventional grids with unidirectional power flow, modern grids operate bidirectionally, so grid authorities are responsible for setting active and reactive power points for participants. Consequently, modern grid codes may have categories of requirements, such as mandatory requirements for grid connection and additional (optional) features. As electric power systems decentralize, it becomes imperative that grid codes arise that translate monetary or benefit-oriented terms of different stakeholders into fair technical functional requirements. In a 100\% RES, every stakeholder in the grid, and in particular those with a renewable resource, can contribute to either an improved or a degraded grid reliability.

In transitioning to a 100\% renewable energy-based power system, grid codes can play a crucial role. These codes need to incorporate functions for control, protection, and stability. Smart inverter control and grid-forming inverters are essential for overcoming limitations associated with renewables. Technical requirements should include voltage and frequency support, inertial response, power quality improvement, and active/reactive power control. Compliance testing can ensure that renewable sources meet grid requirements. As the system evolves, grid codes will continue to adapt to ensure safe, stable, and efficient integration of renewables.

\sectionbreak
\section{Existing solutions}
\label{sec:solutions}
\subsection{Converter based} 

The future power grid will be 100 \% Renewable Energy Source (RES) based, and the power electronic converters will serve as an interface between fluctuating energy sources and the grid. In such a contemporary power system dominated by Solar and Wind energy, the integrity and stability of power grids greatly depend on protection solutions in high-voltage AC systems. While effective in conventional power systems, traditional protection schemes may encounter limitations when applied to converter-based power grids. Such limitations mainly involve voltage fluctuations or faults initiated by the Low Voltage Fault Ride Through (LVRT) events, severely disrupting the operation of the protection relay. A converter-based protection solution provides safeguards to the power grid under such disturbances. In such solutions, fault detection and localization are facilitated by sophisticated techniques to swiftly differentiate between an LVRT and other disturbances, and relays are operated accordingly to avoid any inadvertent malfunctions. By leveraging state-of-the-art communication techniques, real-time monitoring, and phasor measurement unit (PMU) data, these solutions accurately locate faults and facilitate rapid response and fault isolation, minimizing downtime and maintaining grid stability. However, this approach requires significant modifications to the traditional relaying, control logic, and supervisory data control systems, which can incur considerable costs and may not be economic. A more desired approach is directing the signal to the converter to stabilize the power grid during abnormal operations. In general, converter-based protection solutions incorporate redundancy and fault-tolerant features to enhance system resilience and ensure continuous operation, even with intermittent renewable generation. The converter-based solutions can be divided into the following:

\begin{enumerate}
\item Auxiliary signals to distinct fault state from normal (e.g., negative sequence current)
\item Grid forming control (e.g., virtual synchronous machine)
\end{enumerate}

\subsubsection{Auxiliary signals} 

In order to ensure reliable and efficient operation of the electricity grid, auxiliary signals are used as an essential tool for identifying fault states from normal operating conditions in power system protection. These auxiliary signals, which provide additional insight into the system's state beyond primary protection signals, are derived from secondary variables such as voltage, current, frequency, and phase angle measurements. Relays and control systems can better identify transient disturbances, system oscillations or actual faults by incorporating auxiliary signals into protection schemes. Increasing the sensitivity and selectivity of the protection relay is a crucial advantage for auxiliary signals. The relay may be able to differentiate the fault conditions from a normal operation of the system in a precise way, thus preventing unnecessary tripping during temporary disturbances or system fluctuations, through analysing the behaviour of those second variables. In particular, in grids with an increased level of renewable energy integration, where intermittent generation patterns may make fault detection more complicated, this selectivity tripping capacity is essential to maintain system stability and minimise power interruptions. In addition, implementing sophisticated fault detection and location algorithms requires additional signals. For example, an impedance-based fault location algorithm uses voltage and current phase angle measurements to calculate the characteristics of a transmission line so that it can be precisely localised between faults.
Consequently, the protection schemes can quickly identify fault locations, facilitate faster service recovery and minimise interruptions by ancillary signals in these algorithms. In addition to fault detection and location, auxiliary signals are essential in facilitating adaptive protection strategies. These strategies include dynamic adjustment of settings and thresholds based on real-time system conditions such as load levels, network topology, and protection generation patterns. Protection systems can optimise performance, enhance coordination of protection devices, and increase system resilience in response to change operational conditions through continuous monitoring of complementary signals and adjusting the protection settings according to these changes. In general, power system protection technology is making substantial progress in integrating ancillary signals into protective schemes. In addition to enhancing the reliability, selectivity and efficiency of protection schemes, auxiliary signals ultimately contribute to overall grid reliability and resilience by providing additional insight into system dynamics and fault behaviour.

\subsubsection{Grid Forming Control} 

Grid-forming converter-based protection schemes are a revolutionary development in electrical power systems, especially in today's grid frameworks characterised by a significant integration of renewable energy sources. Grid-forming converters have a unique ability to autonomously regulate voltage and frequency, effectively taking on the role of grid stabilisers and forming the grid independently of external synchronisation signals, unlike conventional grid-following converters. This distinctive characteristic enables grid-forming converters to keep the grid stable, even when it operates under an island or weak grid conditions, such as during disturbances or when disconnected from the primary grid. To exploit the full potential of this technology and ensure its reliability and resilience, it is vital to implement protection schemes for grid-forming converters according to their specific needs. These protection schemes must be designed to seamlessly adapt to the dynamic behaviour of grid-forming converters, incorporating sophisticated fault detection, isolation, and system restoration strategies. These protection schemes often incorporate advanced algorithms and communication systems, which allow for rapid detection of faults and responses to maintain the stability of the network. Furthermore, grid-forming converters offer innovative fault transport capabilities to continue providing critical power supply when transient disturbances or faults occur without compromising the system's integrity. It is worth mentioning that protection schemes incorporating grid-forming converters can exploit the maximum capacity of generation converters, improving the reliability, resilience, and sustainability of existing energy systems against changing grid dynamics with a high share of renewable energy resources.

\subsection{IED based}

A number of solutions for protection challenges in converter-dominated power systems described in chapter 3 are based on the new IED logic of protection functions. These solutions among others include: Adaptive protection schemes, Time-Domain Protections, Wavelet based Protection, Centralized protection, AI/ML based Protection.

\subsubsection{Adaptive protection schemes}

Proposed in recent studies, adaptive distance relaying aims to dynamically adjust distance protection settings based on the changing conditions of the power system with high penetration of converter-based sources.\cite{30,39,Adaptive} 
By continuously monitoring parameters such as impedance, voltage, and current, adaptive distance relaying can adapt its protection settings to accommodate variations in system configuration, fault conditions, and operating conditions.
Although adaptive protection schemes offer appealing features, they come with several drawbacks. One of them is related to complexity of their algorithms relying on predefined fault scenarios.

\subsubsection{Time-Domain protections}

Recent advancements in technology enable the utilization of hardware featuring high sampling rates, sample data storage capacity, high processing capabilities, as well as enhanced communication functionalities. These developments facilitate the development of a new generation of protection systems also known as time-domain-based. Time-domain protection functions allows enhanced line protection operating times and fault location detection.

Utilizing time-domain principles involving incremental quantities and traveling waves, protection systems typically achieve operating times in the range of milliseconds. This aspect enables them to actuate within the current surge window, making them a promising option for protection of line connected to converter-based power sources.\cite{TimeDomain, TimeDomain2}

\subsubsection{Wavelet-based protections}

Conventional protection is based on phasors and Fast Fourier Transform (FFT) which is commonly used to extract 50 Hz component from the voltages and currents at the IED input. One of the ways to improve selectivity and sensitivity of relays in case of system with high share of converter-based sources is using the wavelet signal processing algorithm which increase the robustness of protection to the faults in converter-dominated system.

Several examples of wavelet-based protection function algorithms are presented in \cite{wavelet, wavelet2}.

\subsubsection{Centralized protections}

An alternative to solve the issue of converter-based sources impact on protection operation is centralized protection concept.

Centralized protection based on the IEC 61850 standard involves the implementation of protective functions in a centralized manner, utilizing the communication protocols and data models defined in the standard. This approach allows for the coordination and supervision of protection devices and functions from a central location, enabling efficient monitoring and control of the power system. \cite{Centralized, Centralized2}

Several methods and algorithms were suggested based on centralized protection concept including dynamic state estimation based protection, pattern recognition based protection and time synchronization based protection and control.\cite{Centralized}

\subsubsection{AI/ML based protections}

One of the directions to tackle the issue of protective relay malfunctioning in converter-dominated grids is application of Machine Learning (ML) or other Artificial Intelligence (AI) algorithms. This method may include the simulation of all possible fault test scenarios and collection of power system parameters database, which is used for training of the model to solve classification task. The use of advanced tools like neural networks to assist distance protection function is suggested in \cite{AI}.  The results of this approach are promising, even though this solution is hard to implement in practice since generation of data to train and validate the ML model for each protection device is extremely challenging and time-consuming.

The table \ref{table_Existing_Solutoins} shows some of the state-of-the-art literature and protection solutions for high-voltage AC transmission lines.

{\centering
\begin{center}
\begin{table}[h!]
\caption{Existing Solutions}
\label{table_Existing_Solutoins}
\begin{turn}{90}
\centering
\begin{tabular}{lcccccccccccccccc}
\hline
\textbf{Solution}
& \textbf{\cite{30}} & \textbf{\cite{31}} & \textbf{\cite{32}} & \textbf{\cite{33}} & \textbf{\cite{TimeDomain2}} & \textbf{\cite{35}} & \textbf{\cite{wavelet}} & \textbf{\cite{PMU}} & \textbf{\cite{39}} &\textbf{\cite{40}} &\textbf{\cite{wavelet2}} & \textbf{\cite{Adaptive}} & \textbf{\cite{AI}} & \textbf{\cite{TimeDomain} } & \textbf{\cite{Centralized}} & \textbf{\cite{Centralized2}} \\ 
\hline
\hline
\multicolumn{1}{c}{\textbf{Converter}} &   &  &  & &  & &  &  & &  &  &  &  &  &  &  \\
Aux. signals        &  $\times$ & $\checkmark$ & $\checkmark$ & $\checkmark$ & $\times$ & $\checkmark$ & $\times$ & $\times$ & $\checkmark$ & $\checkmark$ & $\times$ & $\times$ & $\times$ & $\times$ & $\times$ & $\times$ \\
GFM Control     &  $\times$ & $\checkmark$ & $\checkmark$ & $\times$ & $\times$ & $\checkmark$ & $\times$ & $\times$ & $\times$\ & $\times$ & $\times$ & $\times$ & $\times$ & $\times$ & $\times$ & $\times$ \\
\hline
\multicolumn{1}{c}{\textbf{IED}} &   &  &  & &  & &  &  & &  &  &  &  &  &  &  \\
Adaptive      &  $\checkmark$ & $\times$ & $\times$ & $\times$ & $\times$ & $\times$ & $\times$ & $\times$ & $\checkmark$ & $\times$ & $\times$ & $\checkmark$ & $\times$ & $\times$ & $\times$ & $\times$ \\
Time-Domain   &  $\times$ & $\times$ & $\times$ & $\times$ & $\checkmark$ & $\times$ & $\times$ & $\checkmark$ & $\times$ & $\times$ & $\times$ & $\times$ & $\times$ & $\checkmark$ & $\times$ & $\times$ \\
Wavelet based            &  $\times$ & $\times$ & $\times$ & $\times$ & $\times$ & $\times$ &  $\checkmark$ & $\times$ & $\times$ & $\times$ & $\checkmark$ & $\times$ & $\times$ & $\times$ & $\times$ & $\times$ \\
PMU based                &  $\times$ & $\times$ & $\times$ & $\times$ & $\times$ & $\times$ &  $\times$ & $\checkmark$ & $\times$ & $\times$ & $\times$ & $\times$ & $\times$ & $\times$ & $\times$ & $\times$ \\
Centralised   &  $\times$ & $\times$ & $\times$ & $\times$ & $\times$ & $\times$ &  $\times$ & $\times$ & $\times$ & $\times$ & $\times$ & $\times$ & $\times$ & $\times$ & $\checkmark$ & $\checkmark$ \\
AI/ML based   &  $\times$ & $\times$ & $\times$ & $\times$ & $\times$ & $\times$ &  $\times$ & $\times$ & $\times$ & $\times$ & $\times$ & $\times$ & $\checkmark$ & $\times$ & $\times$ & $\times$ \\
\hline
\end{tabular}
\end{turn}
\end{table}
\end{center}
}

\sectionbreak
\section{Conclusion}

In conclusion, the transition towards 100\% Renewable Energy Sources (RES) poses unique challenges to power system protection. Low inertia and system strength, coupled with increase of converter-based energy sources, require innovative solutions for effective protection. However, existing protection schemes face numerous challenges, including limitations in distance protection, negative sequence-based protection, and converter control strategies.

Despite these challenges, advancements in technology, such as the introduction of merging units and improvements in computational capabilities and communication infrastructure, present opportunities for centralized protection approaches. The adoption of standards like IEC 61850 further facilitates the implementation of centralized protection schemes, enabling efficient supervision of protective functions and self-healing against hidden failures.

Existing solutions, ranging from converter-based approaches to IED-based methods, offer potential strategies for addressing protection challenges in converter-dominated systems. These include adaptive protection schemes, time-domain protections, wavelet-based protections, centralized protections, and AI/ML-based protections.

In summary, while the transition to 100 \% RES pose a number of challenges for power system protection, it also opens up new avenues for innovation and development.

\sectionbreak
 \printbibliography

@misc{paris_agreement,
   title = {Adoption of the paris agreement - Paris Agreement text English}
}

@misc{cellar,
   author = {European Commission},
   title = {The European Green Deal},
   year = {2019},
}

@article{VANHERTEM2010,
title = {Multi-terminal VSC HVDC for the European supergrid: Obstacles},
journal = {Renewable and Sustainable Energy Reviews},
volume = {14},
number = {9},
pages = {3156-3163},
year = {2010},
issn = {1364-0321},
doi = {https://doi.org/10.1016/j.rser.2010.07.068},
url = {https://www.sciencedirect.com/science/article/pii/S1364032110002480},
author = {Dirk {Van Hertem} and Mehrdad Ghandhari},

}

@book{VANHERTEM2016hvdc,
  title={HVDC Grids: For Offshore and Supergrid of the Future},
  author={Van Hertem, D. and Gomis-Bellmunt, O. and Liang, J.},
  isbn={9781118859155},
  lccn={2016427940},
  series={IEEE Press Series on Power and Energy Systems},
  url={https://books.google.be/books?id=cfWlCgAAQBAJ},
  year={2016},
  publisher={Wiley}
}

@misc{adored,
	author = {European Commission},
	title = {HORIZON EUROPE: Accelerating the Deployment of Offshore wind using Dc technology (ADOreD)},
	howpublished = {\url{https://cordis.europa.eu/project/id/101073554}},
	year = {2022},
	note = {[Accessed 22-02-2024]},
}

@misc{bookprotection,
	author = {Paul M. Anderson},
	title = {{P}ower {S}ystem {P}rotection | {I}{E}{E}{E} e{B}ooks | {I}{E}{E}{E} {X}plore --- ieeexplore.ieee.org},
	howpublished = {\url{https://ieeexplore.ieee.org/servlet/opac?bknumber=5264125}},
	year = {1999},
	note = {[Accessed 29-02-2024]},
}

@phdthesis{orbitjundi,
   author = {Jundi Jia},
   title = {Assessment of Short Circuit Power and Protection Systems for Future Low Inertia Power Systems},
   year = {2018}
}

@article{mukeshpes,
   author = {Mukesh Nagpal and Charles Henville},
   issue = {1},
   journal = {IEEE Transactions on Power Delivery},
   month = {2},
   pages = {62-70},
   publisher = {Institute of Electrical and Electronics Engineers Inc.},
   title = {Impact of Power-Electronic Sources on Transmission Line Ground Fault Protection},
   volume = {33},
   year = {2018},
}

@article{Joule,
title = {The challenges of achieving a 100\% renewable electricity system in the United States},
journal = {Joule},
volume = {5},
number = {6},
pages = {1331-1352},
year = {2021},
issn = {2542-4351},
doi ={https://doi.org/10.1016/j.joule.2021.03.028},
url ={https://www.sciencedirect.com/science/article/pii/S2542435121001513},
author = {Paul Denholm and Douglas J. Arent and Samuel F. Baldwin and Daniel E. Bilello and Gregory L. Brinkman and Jaquelin M. Cochran and Wesley J. Cole and Bethany Frew and Vahan Gevorgian and Jenny Heeter and Bri-Mathias S. Hodge and Benjamin Kroposki and Trieu Mai and Mark J. O’Malley and Bryan Palmintier and Daniel Steinberg and Yingchen Zhang}
}

@misc{irena,
	author = {IRENA},
	title = {Renewable energy statistics 2023},
	year = {2023},
	note = {International Renewable Energy Agency, Abu Dhabi},
}

@misc{IPCC,
	title = {IPCC Special Report on Renewable Energy Sources and Climate Change Mitigation},
	year = {2011}
}

@article{futuresupply,
author = {Armaroli, Nicola and Balzani, Vincenzo},
title = {The Future of Energy Supply: Challenges and Opportunities},
journal = {Angewandte Chemie International Edition},
volume = {46},
number = {1-2},
pages = {52-66},
doi = {https://doi.org/10.1002/anie.200602373},
year = {2007}
}

@misc{EUagency,
author={European Environment Agency},
	title = {Renewable energy in Europe: key for climate objectives, but air pollution needs attention},
	year = {2019}
}

@article{socialacceptance,
title = {Social acceptance of renewable energy innovation: An introduction to the concept},
journal = {Energy Policy},
volume = {35},
number = {5},
pages = {2683-2691},
year = {2007},
issn = {0301-4215},
doi = {https://doi.org/10.1016/j.enpol.2006.12.001},
author = {Rolf Wüstenhagen and Maarten Wolsink and Mary Jean Bürer}
}

@misc{IEA,
	author = {International Energy Agency (IEA)},
	title = {Renewable Energy Technology Roadmap: More than Just Talk},
	howpublished = {\url{https://www.iea.org/news/the-world-is-entering-a-new-age-of-clean-technology-manufacturing-and-countries-industrial-strategies-will-be-key-to-success}},
	year = {2023},
	note = {[Accessed 01-04-2024]},
}

@INPROCEEDINGS{adaptative100,
  author={Patel, Trupal and Brahma, Sukumar and Hernandez-Alvidrez, Javier and Reno, Matthew J.},
  booktitle={2020 IEEE Kansas Power and Energy Conference (KPEC)}, 
  title={Adaptive Protection Scheme for a Real-World Microgrid with 100\% Inverter-Based Resources}, 
  year={2020},
  volume={},
  number={},
  pages={1-6},
  doi={10.1109/KPEC47870.2020.9167527}}

@article{protectionschemes100,
title = {Design Protection Schemes for 100\% Renewable Microgrids},
author = {Wang, JIng and Zamzam, Ahmed and Soumitra, Kumar},
doi = {},
url = {https://www.osti.gov/biblio/1906858}, journal = {},
number = {},
volume = {},
place = {United States},
year = {2022},
month = {12}
}

@ARTICLE{operational100,
  author={Muenz, Ulrich and Xue, Nan and Wu, Xiaofan and Mesanovic, Amer and Cockcroft, Leland and Dangelmaier, Lisa and Heyde, Chris and Dong, Zerui and Chakraborty, Sudipta and Ashok, Aditya},
  journal={IEEE Electrification Magazine}, 
  title={Operational Security in 100\% Inverter-Based Power Systems: Experiences from Hawai‘i}, 
  year={2022},
  volume={10},
  number={3},
  pages={10-19},
  doi={10.1109/MELE.2022.3187632}}

@article{addressingtech,
author = {Hodge, Bri-Mathias S. and Jain, Himanshu and Brancucci, Carlo and Seo, Gab-Su and Korpås, Magnus and Kiviluoma, Juha and Holttinen, Hannele and Smith, James Charles and Orths, Antje and Estanqueiro, Ana and Söder, Lennart and Flynn, Damian and Vrana, Til Kristian and Kenyon, Rick Wallace and Kroposki, Benjamin},
title = {Addressing technical challenges in 100\% variable inverter-based renewable energy power systems},
journal = {WIREs Energy and Environment},
volume = {9},
number = {5},
pages = {e376},
doi = {https://doi.org/10.1002/wene.376},
url = {https://wires.onlinelibrary.wiley.com/doi/abs/10.1002/wene.376},
year = {2020}
}

@ARTICLE{citcea,
  author={Collados-Rodriguez, Carlos and Cheah-Mane, Marc and Prieto-Araujo, Eduardo and Gomis-Bellmunt, Oriol},
  journal={IEEE Transactions on Power Delivery}, 
  title={Stability Analysis of Systems With High VSC Penetration: Where Is the Limit?}, 
  year={2020},
  volume={35},
  number={4},
  pages={2021-2031},
  doi={10.1109/TPWRD.2019.2959541}}

@misc{6,
   author = {Kamala Sarojini Ratnam and K. Palanisamy and Guangya Yang},
   doi = {10.1016/j.rser.2020.109773},
   issn = {18790690},
   journal = {Renewable and Sustainable Energy Reviews},
   month = {5},
   publisher = {Elsevier Ltd},
   title = {Future low-inertia power systems: Requirements, issues, and solutions - A review},
   volume = {124},
   year = {2020},
}

@article{7,
   author = {Qiteng Hong and Md Asif Uddin Khan and Callum Henderson and Agustí Egea-Àlvarez and Dimitrios Tzelepis and Campbell Booth},
   doi = {10.1016/j.eng.2021.06.005},
   issn = {20958099},
   issue = {8},
   journal = {Engineering},
   month = {8},
   pages = {1057-1063},
   publisher = {Elsevier Ltd},
   title = {Addressing Frequency Control Challenges in Future Low-Inertia Power Systems: A Great Britain Perspective},
   volume = {7},
   year = {2021},
}

@article{8,
   author = {Changjun He and Xiuqiang He and Hua Geng and Huadong Sun and Shiyun Xu},
   doi = {10.1109/TEC.2022.3185623},
   issn = {15580059},
   issue = {4},
   journal = {IEEE Transactions on Energy Conversion},
   month = {12},
   pages = {2903-2912},
   publisher = {Institute of Electrical and Electronics Engineers Inc.},
   title = {Transient Stability of Low-Inertia Power Systems With Inverter-Based Generation},
   volume = {37},
   year = {2022},
}

@article{9,
   author = {Xiongfei Wang and Mads Graungaard Taul and Heng Wu and Yicheng Liao and Frede Blaabjerg and Lennart Harnefors},
   doi = {10.1109/OJIA.2020.3020392},
   issn = {26441241},
   journal = {IEEE Open Journal of Industry Applications},
   pages = {115-134},
   publisher = {Institute of Electrical and Electronics Engineers Inc.},
   title = {Grid-Synchronization Stability of Converter-Based Resources - An Overview},
   volume = {1},
   year = {2020},
}

@book{10,
	author = {P. Kundur},
	title = {Power System Stability and Control},
	year = {1994},
	note = {[Accessed 29-02-2024]}
}

@inproceedings{11,
   author = {Helge Urdal and Richard Ierna and Jiebei Zhu and Chavdar Ivanov and Amir Dahresobh and Djaved Rostom},
   doi = {10.1049/iet-rpg.2014.0199},
   issn = {17521424},
   issue = {1},
   journal = {IET Renewable Power Generation},
   month = {1},
   pages = {10-17},
   publisher = {Institution of Engineering and Technology},
   title = {System strength considerations in a converter dominated power system},
   volume = {9},
   year = {2015}
}

@misc{12,
   author = {Md Ohirul Qays and Iftekhar Ahmad and Daryoush Habibi and Asma Aziz and Thair Mahmoud},
   doi = {10.1016/j.rser.2023.113447},
   issn = {18790690},
   journal = {Renewable and Sustainable Energy Reviews},
   month = {9},
   publisher = {Elsevier Ltd},
   title = {System strength shortfall challenges for renewable energy-based power systems: A review},
   volume = {183},
   year = {2023},
}

@incollection{pv,
editor = {L. {Ashok Kumar} and S. {Albert Alexander} and Madhuvanthani Rajendran},
booktitle = {Power Electronic Converters for Solar Photovoltaic Systems},
publisher = {Academic Press},
year = {2021},
isbn = {978-0-12-822730-5},
doi = {https://doi.org/10.1016/B978-0-12-822730-5.04001-0}
}

@INPROCEEDINGS{icrera,
  author={Choi, U. M. and Lee, K. B. and Blaabjerg, F.},
  booktitle={2012 International Conference on Renewable Energy Research and Applications (ICRERA)}, 
  title={Power electronics for renewable energy systems: Wind turbine and photovoltaic systems}, 
  year={2012},
  volume={},
  number={},
  pages={1-8},
  doi={10.1109/ICRERA.2012.6477249}}

@Article{ocean,
AUTHOR = {Garcia-Reyes, Luis A. and Beltrán-Telles, Aurelio and Bañuelos-Ruedas, Francisco and Reta-Hernández, Manuel and Ramírez-Arredondo, Juan M. and Silva-Casarín, Rodolfo},
TITLE = {Level-Shift PWM Control of a Single-Phase Full H-Bridge Inverter for Grid Interconnection, Applied to Ocean Current Power Generation},
JOURNAL = {Energies},
VOLUME = {15},
YEAR = {2022},
NUMBER = {5},
ARTICLE-NUMBER = {1644},
ISSN = {1996-1073},
DOI = {10.3390/en15051644}
}

@article{hydro,
title = {Power electronics in hydro electric energy systems – A review},
journal = {Renewable and Sustainable Energy Reviews},
volume = {32},
pages = {944-959},
year = {2014},
issn = {1364-0321},
author = {R. {Raja Singh} and Thanga {Raj Chelliah} and Pramod Agarwal}
}

@article{DTUwind,
author = {Kaushik Das and Anca D Hansen and Poul E Sørensen},
title ={Understanding IEC standard wind turbine models using SimPowerSystems},
journal = {Wind Engineering},
volume = {40},
number = {3},
pages = {212-227},
year = {2016}
}

@Article{WTenergies,
AUTHOR = {Honrubia-Escribano, Andrés and Jiménez-Buendía, Francisco and Gómez-Lázaro, Emilio and Fortmann, Jens},
TITLE = {Validation of Generic Models for Variable Speed Operation Wind Turbines Following the Recent Guidelines Issued by IEC 61400-27},
JOURNAL = {Energies},
VOLUME = {9},
YEAR = {2016},
NUMBER = {12},
ARTICLE-NUMBER = {1048}
}

@INPROCEEDINGS{WTdynamics,
  author={Ellis, A. and Kazachkov, Y. and Muljadi, E. and Pourbeik, P. and Sanchez-Gasca, J. J.},
  booktitle={2011 IEEE/PES Power Systems Conference and Exposition}, 
  title={Description and technical specifications for generic WTG models — A status report}, 
  year={2011},
  volume={},
  number={},
  pages={1-8}
}

@article{WTcomparative,
title = {Wind Energy Conversion System Topologies and Converters: Comparative Review},
journal = {Energy Procedia},
volume = {162},
pages = {38-47},
year = {2019},
note = {Emerging and Renewable Energy: Generation and Automation},
issn = {1876-6102},
doi = {https://doi.org/10.1016/j.egypro.2019.04.005},
url = {https://www.sciencedirect.com/science/article/pii/S1876610219313645},
author = {Kotb B. Tawfiq and Arafa S. Mansour and Haitham S. Ramadan and Mohamed Becherif and E.E. El-kholy}
}

@misc{DirectIED,
   author = {Alexander Tsylin and Rasmus Kruse-Nielsen and Guangya Yang and Behzad Sedaghat},
   title = {Influence of collection network parameters on performance of distance protection directional elements in offshore wind farms},
   year = {2020},
}

@article{NegSeqIED,
   author = {Aboutaleb Haddadi and Mingxuan Zhao and Ilhan Kocar and Ulas Karaagac and Ka Wing Chan and Evangelos Farantatos},
   doi = {10.1109/TPWRD.2020.2978075},
   issn = {19374208},
   issue = {1},
   journal = {IEEE Transactions on Power Delivery},
   month = {2},
   pages = {289-298},
   publisher = {Institute of Electrical and Electronics Engineers Inc.},
   title = {Impact of inverter-based resources on negative sequence quantities-based protection elements},
   volume = {36},
   year = {2021},
}

@article{SwingIED,
   author = {Chidiebere Okeke and Othman Mohammad Lutfi and Hizam Hashim and Mohd Zainal Abdin Ab Kadir and Noor Izzri Abdul Wahab and Osaji Emmanuel and Samuel Nwagbara and Collins Chimeleze},
   doi = {10.47836/pjst.31.6.14},
   issn = {22318526},
   issue = {6},
   journal = {Pertanika Journal of Science and Technology},
   month = {10},
   pages = {2887-2913},
   publisher = {Universiti Putra Malaysia Press},
   title = {Renewable Energy Impact on Distance Relay Power Swing Blocking and Fault Discrimination: A Review},
   volume = {31},
   year = {2023},
}

@article{CurrentIED,
   author = {Jundi Jia and Guangya Yang and Arne Hejde Nielsen and Peter Roenne‐Hansen},
   doi = {10.1049/joe.2018.0248},
   issn = {2051-3305},
   issue = {15},
   journal = {The Journal of Engineering},
   month = {10},
   pages = {824-829},
   publisher = {Institution of Engineering and Technology (IET)},
   title = {Hardware‐in‐the‐loop tests on distance protection considering VSC fault‐ride‐through control strategies},
   volume = {2018},
   year = {2018},
}

@article{CurImpIED,
   title = {Protection Function Assessment of Present Relays For Wind Generator Applications},
   author = {Jose J. Chavez and Marjan Popov and Alexander Novikov and Sadegh Azizi and Vladimir Terzija},
   journal = {IPST},
   year = {2019},
}

@article{CurImpIED2,
   author = {Ali Hooshyar and Maher A. Azzouz and Ehab F. El-Saadany},
   doi = {10.1109/TPWRD.2014.2369479},
   issn = {08858977},
   issue = {4},
   journal = {IEEE Transactions on Power Delivery},
   month = {8},
   pages = {1770-1780},
   publisher = {Institute of Electrical and Electronics Engineers Inc.},
   title = {Distance Protection of Lines Emanating from Full-Scale Converter-Interfaced Renewable Energy Power Plants-Part I: Problem Statement},
   volume = {30},
   year = {2015},
}

@article{SwingIED2,
   author = {Aboutaleb Haddadi and Ilhan Kocar and Ulas Karaagac and Henry Gras and Evangelos Farantatos},
   doi = {10.1109/TPWRD.2019.2896135},
   issn = {19374208},
   issue = {3},
   journal = {IEEE Transactions on Power Delivery},
   month = {6},
   pages = {1118-1128},
   publisher = {Institute of Electrical and Electronics Engineers Inc.},
   title = {Impact of Wind Generation on Power Swing Protection},
   volume = {34},
   year = {2019},
}

@article{PhaseIED,
   author = {Ritwik Chowdhury and Evangelos Farantatos},
   doi = {10.13140/RG.2.2.22715.90406},
   title = {Protection Challenges and Practices for Interconnecting Inverter Based Resources to Utility Transmission Systems},
   url = {https://www.researchgate.net/publication/364717442},
   year = {2020}
}

@article{OverviewIED,
   author = {Aboutaleb Haddadi and Evangelos Farantatos and Ilhan Kocar and Ulas Karaagac},
   doi = {10.3390/en14041050},
   issn = {19961073},
   issue = {4},
   journal = {Energies},
   month = {2},
   publisher = {MDPI AG},
   title = {Impact of inverter based resources on system protection},
   volume = {14},
   year = {2021},
}

@article{rocofIED,
   author = {Aboutaleb Haddadi and Evangelos Farantatos and Ilhan Kocar and Ulas Karaagac},
   title = {Implications of Reduced Inertia Levels on the Electricity System Technical Report on the Challenges and Solutions for System Operations with Very High Penetrations of Non-Synchronous Resources},
   url = {www.epri.com}
}

@misc{entsoeIED,
        author = {ENTSOE},
	title = {Inertia and Rate of Change of Frequency},
	year = {2020},
	note = {[Accessed 23-03-2024]},
}

@article{30,
  title={Distance protection for lines connecting converter interfaced renewable power plants: adaptive to grid-end structural changes},
  author={George, Neethu and Naidu, OD and Pradhan, Ashok Kumar},
  journal={IEEE Transactions on Power Delivery},
  year={2022},
  publisher={IEEE}
}

@article{31,
  title={Active phase control to enhance distance relay in converter-interfaced renewable energy systems},
  author={Zhang, Haobo and Xiang, Wang and Hong, Qiteng and Wen, Jinyu},
  journal={International Journal of Electrical Power \& Energy Systems},
  volume={143},
  pages={108433},
  year={2022},
  publisher={Elsevier}
}

@article{32,
  title={A power-angle-based adaptive overcurrent protection scheme for grid-forming inverter under large grid disturbances},
  author={Huang, Liang and Wu, Chao and Zhou, Dao and Blaabjerg, Frede},
  journal={IEEE Transactions on Industrial Electronics},
  volume={70},
  number={6},
  pages={5927--5936},
  year={2022},
  publisher={IEEE}
}

@article{33,
  title={A control method for converter-interfaced sources to improve operation of directional protection elements},
  author={Yang, Zhe and Liu, Zhou and Zhang, Qi and Chen, Zhe and de Jesus Chavez, Jose and Popov, Marjan},
  journal={IEEE Transactions on Power Delivery},
  volume={38},
  number={1},
  pages={642--654},
  year={2022},
  publisher={IEEE}
}

@article{TimeDomain2,
  title={A novel directional element for transmission line connecting inverter-interfaced renewable energy power plant},
  author={Wang, Yuxi and Wen, Minghao and Chen, Yu},
  journal={International Journal of Electrical Power \& Energy Systems},
  volume={145},
  pages={108682},
  year={2023},
  publisher={Elsevier}
}

@article{35,
  title={A control-based solution for distance protection of lines connected to converter-interfaced sources during asymmetrical faults},
  author={Banaiemoqadam, Amin and Hooshyar, Ali and Azzouz, Maher A},
  journal={IEEE Transactions on Power Delivery},
  volume={35},
  number={3},
  pages={1455--1466},
  year={2019},
  publisher={IEEE}
}

@article{wavelet,
  title={A novel structural similarity based pilot protection for renewable power transmission line},
  author={Zheng, Liming and Jia, Ke and Bi, Tianshu and Yang, Zhe and Fang, Yu},
  journal={IEEE Transactions on Power Delivery},
  volume={35},
  number={6},
  pages={2672--2681},
  year={2020},
  publisher={IEEE}
}

@article{PMU,
  title={Back up protection scheme for high impedance faults detection in transmission systems based on synchrophasor measurements},
  author={Vlahini{\'c}, Sa{\v{s}}a and Frankovi{\'c}, Dubravko and Juri{\v{s}}a, Borna and Zbunjak, Zoran},
  journal={IEEE Transactions on Smart Grid},
  volume={12},
  number={2},
  pages={1736--1746},
  year={2020},
  publisher={IEEE}
}

@article{39,
  title={Adaptive distance protection for multi-terminal lines connecting converter-interfaced renewable energy sources},
  author={Paladhi, Subhadeep and Hong, Qiteng and Booth, Campbell D},
  year={2022},
  publisher={IET}
}

@article{40,
  title={Protection collaborative fault control for power electronic-based power plants during unbalanced grid faults},
  author={Ma, Kaiqi and Chen, Zhe and Liu, Zhou and Bak, Claus Leth and Castillo, Manuel},
  journal={International Journal of Electrical Power \& Energy Systems},
  volume={130},
  pages={107009},
  year={2021},
  publisher={Elsevier}
}

@article{wavelet2,
  title={Ultra-high-speed transient-based directional relay for AC transmission lines connected to LCC-HVDC inverter station},
  author={Hao, Li and Jiandong, Duan and Yang, Lei and Lihao, Tuo},
  journal={International Journal of Electrical Power \& Energy Systems},
  volume={123},
  pages={106235},
  year={2020},
  publisher={Elsevier}
}

@article{43,
author = {Marchgraber, Jürgen and Gawlik, Wolfgang},
year = {2020},
month = {05},
pages = {2484},
title = {Dynamic Voltage Support of Converters during Grid Faults in Accordance with National Grid Code Requirements},
volume = {13},
journal = {Energies},
doi = {10.3390/en13102484}
}

@INPROCEEDINGS{44,
  author={Mohanty, Rabindra and Chen, Peiyuan and Tuan, Le Anh},
  booktitle={15th International Conference on Developments in Power System Protection (DPSP 2020)}, 
  title={Current restrained undervoltage protection scheme of converter dominated microgrids}, 
  year={2020},
  volume={},
  number={},
  pages={1-6},
  doi={10.1049/cp.2020.0021}}

@INPROCEEDINGS{45,
  author={Cao, Shuxiu and Hong, Qiteng and Liu, Di and Booth, Campbell},
  booktitle={2023 IEEE Belgrade PowerTech}, 
  title={Review and Evaluation of Control-Based Protection Solutions for Converter-Dominated Power Systems}, 
  year={2023},
  volume={},
  number={},
  pages={01-07},
  doi={10.1109/PowerTech55446.2023.10202927}}

@ARTICLE{46,
  author={Liu, Teng and Wang, Xiongfei and Liu, Fangcheng and Xin, Kai and Liu, Yunfeng},
  journal={IEEE Open Journal of Power Electronics}, 
  title={Transient Stability Analysis for Grid-Forming Inverters Transitioning from Islanded to Grid-Connected Mode}, 
  year={2022},
  volume={3},
  number={},
  pages={419-432},
  doi={10.1109/OJPEL.2022.3189801}}

@ARTICLE{47,
  author={Huang, Linbin and Xin, Huanhai and Wang, Zhen and Zhang, Leiqi and Wu, Kuayu and Hu, Jiabing},
  journal={IEEE Transactions on Smart Grid}, 
  title={Transient Stability Analysis and Control Design of Droop-Controlled Voltage Source Converters Considering Current Limitation}, 
  year={2019},
  volume={10},
  number={1},
  pages={578-591},
  doi={10.1109/TSG.2017.2749259}}

@ARTICLE{49,
  author={Qoria, Taoufik and Gruson, François and Colas, Fréderic and Denis, Guillaume and Prevost, Thibault and Guillaud, Xavier},
  journal={IEEE Journal of Emerging and Selected Topics in Power Electronics}, 
  title={Critical Clearing Time Determination and Enhancement of Grid-Forming Converters Embedding Virtual Impedance as Current Limitation Algorithm}, 
  year={2020},
  volume={8},
  number={2},
  pages={1050-1061},
  doi={10.1109/JESTPE.2019.2959085}}

@article{50,
title = {Current limiting algorithms and transient stability analysis of grid-forming VSCs},
journal = {Electric Power Systems Research},
volume = {189},
pages = {106726},
year = {2020},
issn = {0378-7796},
doi = {https://doi.org/10.1016/j.epsr.2020.106726},
url = {https://www.sciencedirect.com/science/article/pii/S0378779620305290},
author = {Taoufik Qoria and François Gruson and Fréderic Colas and Xavier Kestelyn and Xavier Guillaud}
}

@INPROCEEDINGS{51,
  author={Yang, Kexin and Bao, Yanhong and Ren, Xiancheng and Wu, Feng and Zhang, Jinlong},
  booktitle={2023 6th Asia Conference on Energy and Electrical Engineering (ACEEE)}, 
  title={Influence Analysis of Control Parameters on Transient Stability of Single-loop Voltage-Magnitude Controlled Grid-Forming Converter}, 
  year={2023},
  volume={},
  number={},
  pages={17-24},
  doi={10.1109/ACEEE58657.2023.10239584}}

@INPROCEEDINGS{52,
  author={Ge, Pingjuan and Tu, Chunming and Xiao, Fan and Xiao, Zekun and Hou, Yuchao and Chen, Leqi},
  booktitle={2022 International Conference on Power Energy Systems and Applications (ICoPESA)}, 
  title={Transient Stability Boundary of a Droop-Controlled Voltage Source Converter with Different Control Parameters}, 
  year={2022},
  volume={},
  number={},
  pages={286-291},
  doi={10.1109/ICoPESA54515.2022.9754460}}

@inproceedings{Adaptive,
   author = {Amila Pathirana and Athula Rajapakse and Adeyemi Charles Adewole},
   title = {Implementation and Testing of a Hybrid Protection Scheme for Active Distribution Network},
   publisher = {2018 IEEE Electrical Power and Energy Conference (EPEC).},
   isbn = {9781538654194}
}

@inproceedings{AI,
   author = {Chhavi Lal and S. Sarangi and S. R. Mohanty and Asheesh K. Singh},
   doi = {10.1109/UPCON50219.2020.9376461},
   isbn = {9780738111513},
   journal = {7th IEEE Uttar Pradesh Section International Conference on Electrical, Electronics and Computer Engineering, UPCON 2020},
   month = {11},
   publisher = {Institute of Electrical and Electronics Engineers Inc.},
   title = {ANN Based Adaptive Mho Distance Protection in Distribution Network with Distributed Generations},
   year = {2020}
}

@misc{TimeDomain,
   author = {Ricardo Abboud and David Dolezilek},
   title = {Time-Domain Technology-Benefits to Protection, Control, and Monitoring of Power Systems Automation for Electric Power Systems, April 2017, under the title "Case Study of Time-Domain Protection: Field-Proven Benefits to Protection, Control, and Monitoring" Time-Domain Technology-Benefits to Protection, Control, and Monitoring of Power Systems},
}

@misc{Centralized,
   author = {M Adamiak and J Jester and Y Luskind and J Park and A Apostolov and M Kezunovic and V Madani and C Preuss and G Antonova and M Kockott and A P Meliopoulos and Q Qiu and S Brahma and L Kojovic and R Midence and M Sachdev and M Dadashzadeh and R Lascu and P Myrda and I Voloh and R Hunt and Y Liao and A Oliveira and J Xavier and Michael Thompson and Martin Best and William English and Vahid Madani and Lubomir Sevov and Gustavo Brunello and Dominick Fontana and Dean Miller and Charles Sufana and Patrick Carroll and Stephen Grier and George Moskos and Qun Qiu and Arvind Chaudhary and Randy Hamilton and Chuck Mozina and Ilia Voloh and Stephen Conrad and Roger Hedding and Adi Mulawarman and John Wang and Randy Crellin and Gene Henneberg and Pratap Mysore and Roger Whittaker Paul and Elkin Charles and Henville Mukesh and Nagpal Rich and Young Gerald and Johnson Bruce Pickett},
   title = {2015-December WG K15 Report-Centralized Substation Protection and Control i Centralized Substation Protection and Control IEEE PES Power System Relaying Committee Members},
   year = {2015}
}

@article{Centralized2,
   author = {Hussain F. Albinali and A. P. Sakis Meliopoulos},
   doi = {10.1109/TPWRD.2017.2789318},
   issn = {08858977},
   issue = {3},
   journal = {IEEE Transactions on Power Delivery},
   month = {6},
   pages = {1418-1427},
   publisher = {Institute of Electrical and Electronics Engineers Inc.},
   title = {Resilient Protection System Through Centralized Substation Protection},
   volume = {33},
   year = {2018},
}

@article{international2021assessing,
  title={Assessing the effects of economic recoveries on global energy demand and CO2 emissions in 2021},
  author={International Energy Agency},
  journal={Global Energy Rev},
  year={2021}
}

@incollection{sathiyanarayanan2024grid,
  title={Grid interconnection standards, grid code requirements and compliance for renewable integration},
  author={Sathiyanarayanan, T and others},
  booktitle={Power Systems Operation with 100\% Renewable Energy Sources},
  pages={55--64},
  year={2024},
  publisher={Elsevier}
}

@article{tsili2009review,
  title={A review of grid code technical requirements for wind farms},
  author={Tsili, Marina and Papathanassiou, S},
  journal={IET Renewable power generation},
  volume={3},
  number={3},
  pages={308--332},
  year={2009},
  publisher={IET}
}

@article{khalid2022efficient,
  title={Efficient load frequency control of renewable integrated power system: A twin delayed DDPG-based deep reinforcement learning approach},
  author={Khalid, Junaid and Ramli, Makbul AM and Khan, Muhammad Saud and Hidayat, Taufal},
  journal={IEEE Access},
  volume={10},
  pages={51561--51574},
  year={2022},
  publisher={IEEE}
}

@inproceedings{erlich2005grid,
  title={Grid code requirements concerning connection and operation of wind turbines in Germany},
  author={Erlich, Istv{\'a}n and Bachmann, Udo},
  booktitle={IEEE Power Engineering Society General Meeting, 2005},
  pages={1253--1257},
  year={2005},
  organization={Ieee}
}

@misc{irena2022grid,
  title={Grid Codes for Renewable Powered Systems},
  author={IRENA, MS},
  year={2022},
  publisher={International Renewable Energy Agency Abu Dhabi, United Arab Emirates}
}

@INPROCEEDINGS{SCR2,
  author={Damanik, Oscar and Sakinci, Özgür Can and Grdenić, Goran and Beerten, Jef},
  booktitle={2022 IEEE PES Innovative Smart Grid Technologies Conference Europe (ISGT-Europe)}, 
  title={Evaluation of the use of short-circuit ratio as a system strength indicator in converter-dominated power systems}, 
  year={2022},
  volume={},
  number={},
  pages={1-5},
  doi={10.1109/ISGT-Europe54678.2022.9960381}}

@INPROCEEDINGS{SCR3,
  author={Gavrilovic, A.},
  booktitle={International Conference on AC and DC Power Transmission}, 
  title={AC/DC system strength as indicated by short circuit ratios}, 
  year={1991},
  volume={},
  number={},
  pages={27-32},
  doi={}}

@misc{Gordon,
	author = {Samuel GORDON, Qiteng HONG, Keith BELL},
	title = {Implications of Reduced Fault Level and its Relationship to System Strength:  A Scotland Case Study },
	howpublished = {\url{https://pureportal.strath.ac.uk/files/135647782/Gordon_etal_CIGRE_2022_Implications_of_reduced_fault_level_and_its_relationship_to_system_strength.pdf}},
	year = {2022},
	note = {[Accessed 16-07-2024]},
}

@misc{Gordon2,
	author = {Samuel GORDON, ConnorMcGarry, Keith BELL},
	title = {The growth of distributed generation and associated challenges: AGreat Britain case study },
	howpublished = {\url{https://doi.org/10.1049/rpg2.12416}},
	year = {2022},
	note = {[Accessed 16-07-2024]},
}

@INPROCEEDINGS{debates,
  author={Boričić, Aleksandar and Torres, Jose Luis Rueda and Popov, Marjan},
  booktitle={2022 IEEE PES Innovative Smart Grid Technologies - Asia (ISGT Asia)}, 
  title={System Strength: Classification, Evaluation Methods, and Emerging Challenges in IBR-dominated Grids}, 
  year={2022},
  volume={},
  number={},
  pages={185-189},
  doi={10.1109/ISGTAsia54193.2022.10003499}}
\sectionbreak

\end{document}